# Nucleation of titanium nanoparticles in an oxygen-starved environment, II: Theory

**Rickard Gunnarsson[1], Nils Brenning[1, 2*], Lars Ojamäe[1], Emil Kalered[1], Michael Allan Raadu[2], and Ulf Helmersson[1]**

[1] *IFM-Material Science, Linköping University, 581 83 Linköping, Sweden*
[2] *KTH Royal Institute of Technology, EECS, Department of Space and Plasma Physics, SE-100 44, Stockholm, Sweden*
**corresponding author (e-mail: nils.brenning@ee.kth.se)*

**Abstract:** The nucleation and growth of pure titanium nanoparticles in a low-pressure sputter plasma has been believed to be essentially impossible. The addition of impurities, such as oxygen or water, facilitates this and allows the growth of nanoparticles. However, it seems that this route requires so high oxygen densities that metallic nanoparticles in the hexagonal $\alpha$Ti-phase cannot be synthesized. Here we present a model which explains results for the nucleation and growth of titanium nanoparticles in the absent of reactive impurities. In these experiments, a high partial pressure of helium gas was added which increased the cooling rate of the process gas in the region where nucleation occurred. This is important for two reasons. First, a reduced gas temperature enhances $Ti_2$ dimer formation mainly because a lower gas temperature gives a higher gas density, which reduces the dilution of the Ti vapor through diffusion. The same effect can be achieved by increasing the gas pressure. Second, a reduced gas temperature has a "more than exponential" effect in lowering the rate of atom evaporation from the nanoparticles during their growth from a dimer to size where they are thermodynamically stable, $r^*$. We show that this early stage evaporation is not possible to model as a thermodynamical equilibrium. Instead, the single-event nature of the evaporation process has to be considered. This leads, counter intuitively, to an evaporation probability from nanoparticles that is exactly zero below a critical nanoparticle temperature that is size-dependent. Together, the mechanisms described above explain two experimentally found limits for nucleation in an oxygen-free environment. First, there is a lower limit to the pressure for dimer formation. Second, there is an upper limit to the gas temperature above which evaporation makes the further growth to stable nuclei impossible.

## 1. Introduction

It is commonly reported that a supply of small amounts oxygen is necessary for the nucleation and growth of titanium nanoparticles in the gas phase by sputtering techniques [1] [2] [3] [4]. This oxygen can be leaked into the process or originate from contaminants such as $H_2O$. The need for oxygen in these experiments has been attributed to the much higher binding energy in TiO dimers, as compared to $Ti_2$-dimers. This is important since dimer formation is a necessary first step in the formation of nanoparticles. Although adding oxygen can be used to stimulate nucleation, it can also cause problems such as reacting with the target and influencing the particle stoichiometry. In a companion experimental paper [4] we report that the growth of titanium nanoparticles in the metallic hexagonal crystal phase, $\alpha$Ti, is possible by adding a high partial pressure of helium to the process instead of small amounts of oxygen. The subject of the present work is theoretical understanding: both to unravel the role of oxygen in the oxygen-aided nucleation process, and to understand the physics of the nucleation process in the absence of oxygen.

The analysis here is based on results from two experiments, one in a high vacuum system [5] with nucleation in an oxygen-containing environment, and one in a ultra-high vacuum (UHV) system [4]. In the latter, helium replaces oxygen as being necessary for nucleation. Both experiments use a discharge type shown in figure 1 (a). Argon gas is let in through a hollow cathode of Ti, to which short electric pulses with momentary high power and with a low duty cycle are applied. During the pulses Ti atoms are sputtered out from inside the hollow cathode and, to a large degree, ionized in the intense plasma created. After each pulse, a cloud of Ti and $Ti^+$ is ejected out of the hollow cathode. It is in the region outside the hollow cathode the nanoparticles are most likely to nucleate and begin to grow, within a range of distances from the cathode where the two necessary conditions for nucleation are met: a sufficiently low gas temperature, and high enough density of growth material. After growing to their final size they are transported by the gas flow and are collected on the substrate which has a positive electric bias.

The situation is complicated by the fact that the environment in which the nanoparticles nucleate and grow is characterized both by strong gradients and by rapid time variations. Already in the steady state situation, between the pulses, the wall temperature inside the hollow cathode is elevated, and this will heat the Ar gas which is fed through it. This temperature is, for the pulse parameters used herein, estimated to be above 1000 K. In contrast the walls of the external chamber, and the He gas injected into it, are typically kept at room temperature. There is therefore, already in the steady state between pulses, a "mixing zone" outside the hollow cathode orifice in which both the gas temperature and gas composition change. The size of this mixing zone, and the gradients within it, depend on gas flow, pressure, process gas species, and boundary temperatures. During the discharge pulses, this steady state situation is further complicated by temporal variations as the hot cloud containing the sputtered growth material is ejected into the mixing zone. The cloud of growth material is then both convected with the process gas flow, and expands through it by diffusion (atoms) and ambipolar diffusion (ions). It is in this complicated environment that we need to discuss the processes of nucleation and growth.

For the theoretical discussion the regions in the experimental setup are schematically divided up in to three zones defined by the state between pulses. Zone 1 is inside the hollow cathode where it is assumed to be too hot for nanoparticles to nucleate. Zone 2 is where the hot gas that flows out from the hollow cathode is mixed with the colder helium gas. This zone is defined as the region in which the gas temperature and the gas mixture have significant gradients. It is within this zone that the nanoparticles are most likely to nucleate due to a combination of lower temperature than in zone 1, and a high titanium density during the pulses. Zone 3 is defined as the region where the argon gas is well mixed with the helium, and where the gas temperature is the same as the vacuum chamber wall temperature. For details on the experimental arrangements, see Gunnarsson *et al* [4][5].

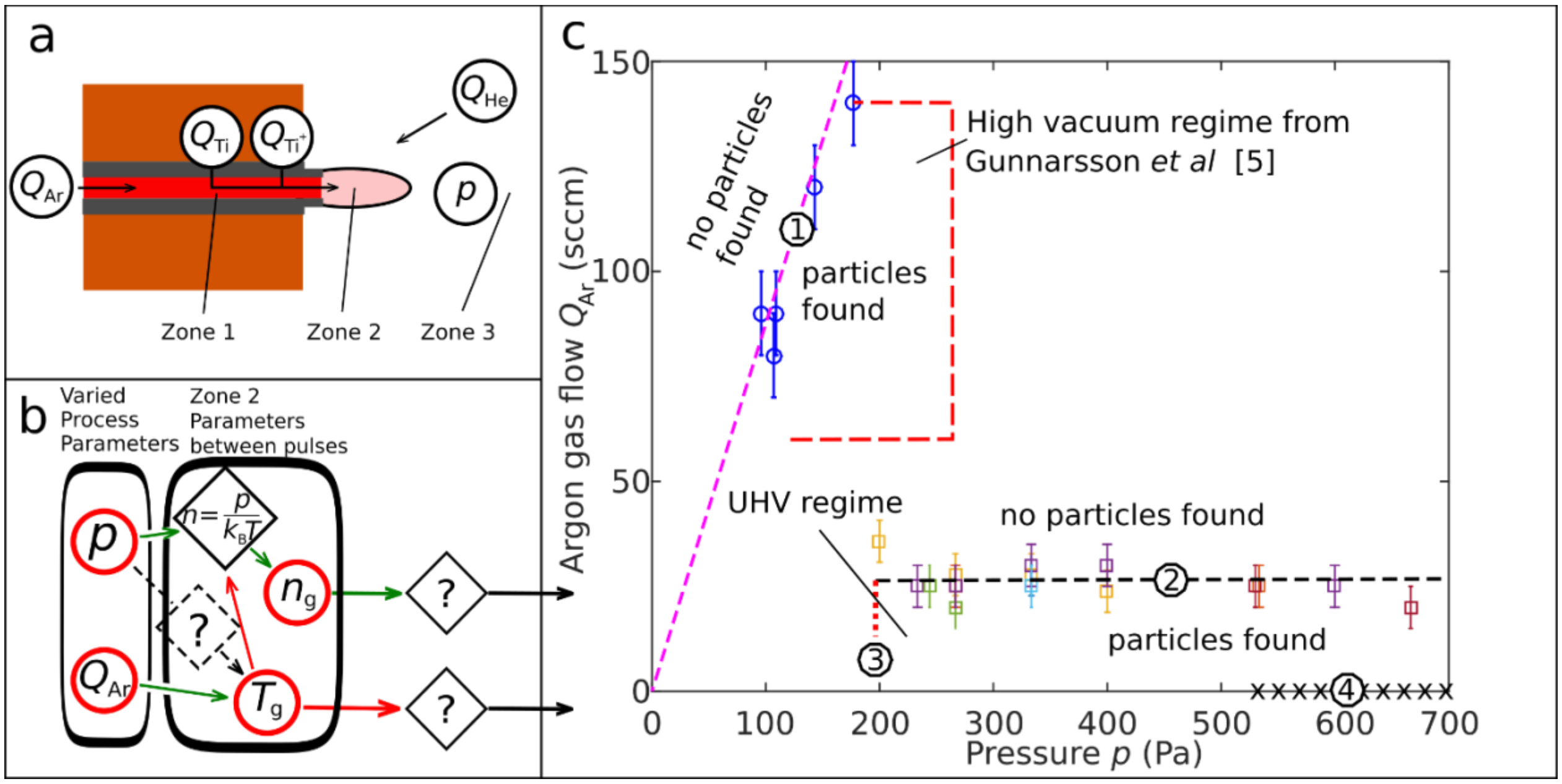

*Figure 1 An overview of how the present study is related to the experimental results from earlier work. (a) The experimental device, and zones 1, 2, and 3 as defined for the time between pulses. The hollow cathode has an inner diameter of 5 mm, the pulse frequency is 1500 Hz, and the pulse length is 80 μs. The pressure in the chamber is controlled independently from the Ar flow rate through a throttle valve. The external process parameters shown are five:* $Q_{Ar}, Q_{He}, p$, $Q_{Ti}$ *and* $Q_{Ti^+}$. *(b) A flow chart for the relations between the two herein varied external process parameters,* $p$ *and* $Q_{Ar}$, *and the two key internal gas parameters* $n_g$ *and* $T_g$ *in zone 2, adapted from* [6]. *Green arrows show when an increase in the parameter/process at the start of the arrow increases the parameter/process at the arrow head, and red arrows show the opposite influence. The theoretical understanding of how* $T_g$ *and* $n_g$ *influence the nucleation process, marked with question marks, is the main subject of the present paper. (c) Four different types of limits for nucleation of nanoparticles, marked (1), (2), (3), (4), as experimentally found in* $(p, Q_{Ar})$ *surveys* [4] [6].

Figure 1 (a) illustrates five process parameters that will be discussed in this paper: the argon gas flow $Q_{\mathrm{Ar}}$, the helium gas flow $Q_{\mathrm{He}}$, the pressure $p$, and the fluxes $Q_{\mathrm{Ti}}$ and $Q_{\mathrm{Ti}^+}$, into zone 1, of growth material. The latter two are determined by the electric pulse parameters. In addition, variations of the wall temperature $T_{\mathrm{wall}}$, and of the addition of an oxygen flow $Q_{\mathrm{O}_2}$ with the helium gas, were explored in the companion paper [4]. These five parameters give a too large parameter space to be fully treated theoretically with reasonable effort. We therefore limit the present work by keeping all parameters fixed except $p$ and $Q_{\mathrm{Ar}}$. These two are chosen because they are found to have a strong combined influence on the nanoparticle formation [4] [5]. We call this type of study a "$(p, Q_{\mathrm{Ar}})$ survey". Such surveys were the key tool in analyzing the final size of nanoparticles as function of pressure presented by Gunnarsson *et al* [5]. Here, and in the companion paper [4] we use $(p, Q_{\mathrm{Ar}})$ surveys for evaluations of the existence of nanoparticles (independent of size), in order to assess under which $(p, Q_{\mathrm{Ar}})$ combinations nanoparticles are created. Our criterion to identify conditions where nanoparticles are generated is that a deposit should be observable by ocular inspection after 10 minutes of exposure to the plasma. This method was found to give reproducible limits in the $(p, Q_{\mathrm{Ar}})$ survey, and is also in close agreement with a more accurate determination by SEM analysis [4]. We here assume that the absence of nanoparticles are due a bottleneck in the nucleation phase, *i.e.,* that neither the subsequent growth of the nuclei, nor the transport of the nanoparticles to the substrate, is the problem. This assumption will be verified *a posteriori* since we show that the observed limits are consistent with a model for the nucleation process. We are investigating two possible bottlenecks in the nanoparticle nucleation: the creation of dimers, and the growth from dimers to a stable size $r^*$ (here defined as the size at which a given nanoparticle is more likely grow further than to shrink by evaporation).

Figure 1 (c) shows four limits to nucleation that were found experimentally in $(p, Q_{\mathrm{Ar}})$ surveys. The limit marked (1) was found in a high vacuum system, and was in [5] identified to occur at a constant ratio $p/Q_{\mathrm{Ar}}$. This was shown to be consistent with a required lowest level of the concentration of the impurities that are always present in high vacuum systems. The final nanoparticles in these experiments always had an oxide crystal phase. The role of the impurities (probably water) was therefore proposed to be that they enabled oxidation of the nanoparticles already during the nucleation stage, giving particles that were more stable against evaporation. We therefore call this oxygen-assisted nucleation, and call the limit marked (1) in the $(p, Q_{\mathrm{Ar}})$ survey the "oxygen limit" for nucleation. In the experiment in [5] the oxygen limit for nucleation had the form

$$\frac{p}{Q_{\mathrm{Ar}}} > 1.3 \text{ [Pa/sccm]} \qquad (1)$$

In the UHV system studied in the experimental companion paper [4], the nanoparticles is produced at such low impurity concentration that they obtain the metallic $\alpha$Ti-phase, far below the oxygen limit for nucleation (1) found in [5]. Here, two other types of limits for nucleation are identified in the $(p, Q_{\mathrm{Ar}})$ survey. Nanoparticles is only found above a pressure limit, the "$p$ limit" for nucleation,

$$p > 200 \text{ [Pa]} \qquad (2)$$

which is marked (3) in figure 1 (c). Above this pressure, nanoparticles is only found below a gas flow limit, the "$Q_{\mathrm{Ar}}$ limit", approximately

$$Q_{\mathrm{Ar}} < 25 \text{ [sccm]}, \qquad (3)$$

which is marked (2) in the figure. It should be noted that the $Q_{\mathrm{Ar}}$ limit varies between experiments, but average at around $Q_{\mathrm{Ar}}$= 20-30 sccm for the full pressure range investigated. There is no clear pressure dependence of the $Q_{\mathrm{Ar}}$ limit.

The limit marked (4),

$$Q_{\mathrm{Ar}} > 0 \tag{4}$$

represents that no nanoparticles are found at zero argon gas flow, *i.e.,* in a pure helium discharge. This limit (marked by crosses in figure 1 (c)) is only drawn in the pressure range above 530 Pa. The reason is that a process instability makes it impossible to operate the discharge at combinations $Q_{\mathrm{Ar}} = 0$ and $Q_{\mathrm{He}} < 530$ Pa.

We will not be able to make a quantitative theory which explains the specific numerical constants in the equations above. We will instead show that the *forms* of these relations are consistent with a proposed set of mechanisms involved in the nucleation. The theoretical approach is illustrated by the two left-hand panels in figure 1. For the time between the pulses, the two investigated process parameters $p$ and $Q_{\mathrm{Ar}}$ determine the situation in the growth zone. Of particular interest is zone 2 immediately outside the hollow cathode. This zone is defined as where the gas temperature gradually changes, from an estimated 1000 – 1500 K inside the hollow cathode [4], to 300 K in zone 3. In this zone there is also a mixture of Ar and He gas. When growth material created in zone 1 is ejected as a cloud into zone 2 it starts to expanding by diffusion through the gas environment. The local gas density $n_{\mathrm{g}}$ and gas temperature $T_{\mathrm{g}}$ in zone 2 will determine if there will be significant nucleation before the growth material has become too diluted. The answer depends on the type of nucleation, oxygen-assisted or without oxygen, and also depends on which stage in nucleation is critical: the dimer formation, or the growth to stable nuclei $r*$.

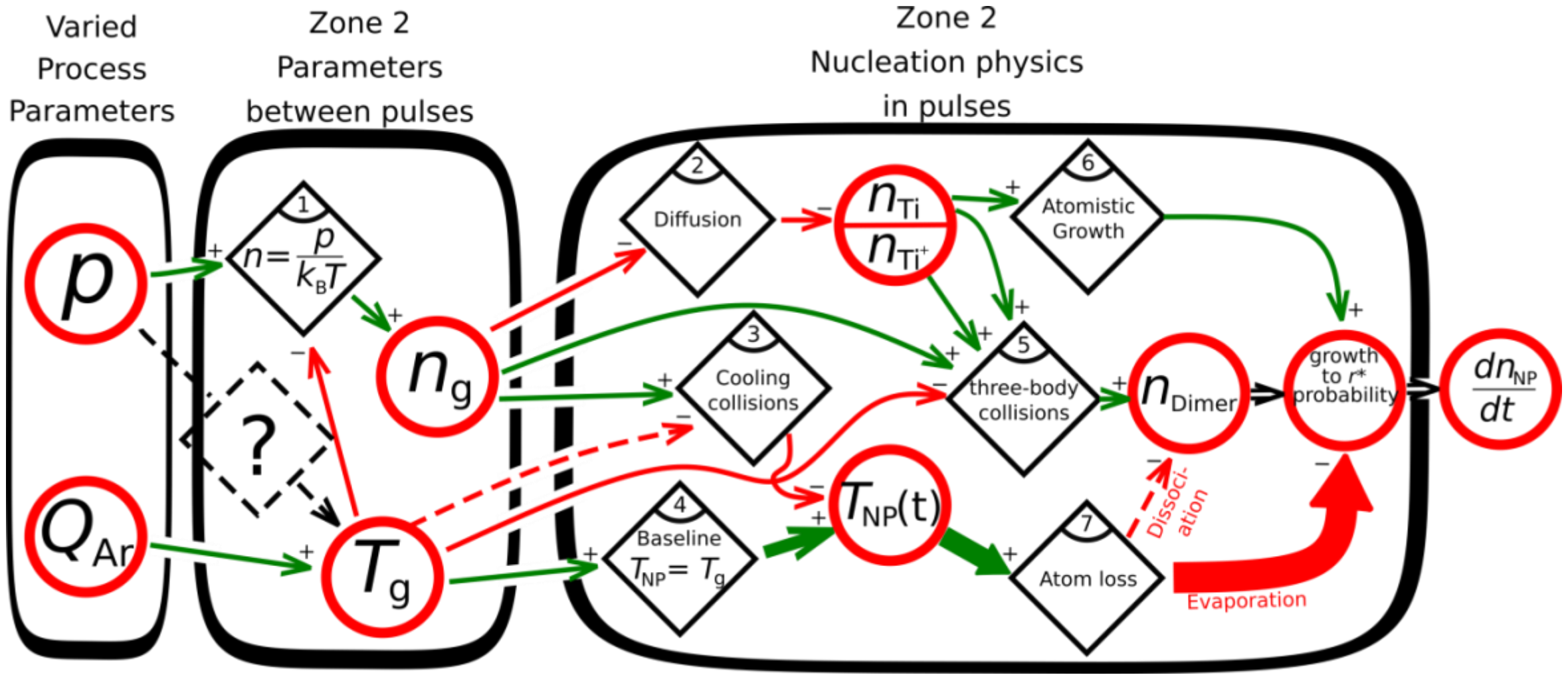

*Figure 2 A process flow chart for the case where p and $Q_{Ar}$ are varied in an oxygen-starved environment. Parameters are drawn in circles, and processes in diamonds. The colors of the arrows denote the sign of the influence, and their thickness indicates the strength of the influence, as described in the text.*

The theoretical questions addressed here are symbolized by the arrows in figure 1 (b): how are the process parameters $p$ and $Q_{\mathrm{Ar}}$ related to $n_{\mathrm{g}}$ and $T_{\mathrm{g}}$ in zone 2, and how are these two parameters in turn related to the experimentally observed nucleation limits shown in figure 1 (c)? The paper is organized as follows. Section 2 contains a one-by-one analysis of individual mechanisms that are involved in the nucleation process, and section 3 contains a discussion which puts these mechanisms into a common context. Section 4, finally, contains a summary and a discussion.

## 2. Processes analyzed one by one

Figure 1 (b) showed a flow chart for the influence, from the two varied process parameters $p$ and $Q_{\mathrm{Ar}}$, to the gas parameters $n_{\mathrm{g}}$ and $T_{\mathrm{g}}$ in zone 2 between the pulses. This flow chart is included to the left in figure 2, a complete flow chart including the whole nucleation process in an oxygen-starved environment. The new processes and parameters in figure 2 correspond to the two question-marked arrows in figure 1 (b), and contain the nucleation physics in the clouds of growth material that are ejected into zone 2.

Parameters are shown in circles, and processes in diamonds. The $+/-$ signs and the colors of the arrows denote the sign of the effect on the parameter (or the process) at the arrowhead, when the parameter (or the process) at the start of the arrow is enhanced. Green$(+)$ denotes increase, and red$(-)$ denotes decrease. The thickness of an arrow indicates the sensitivity of this influence. This type of flow chart is useful in keeping track of the complicated interplay between processes. By following one individual sequence of influences, from the process parameters to the nucleation rate, an even number of red arrows shows a positive influence, while an odd number shows a negative influence. For example: an <u>increase</u> in the gas density $n_{\mathrm{g}}$ (with $T_{\mathrm{g}}$ unchanged) increases the rate of cooling collisions on a nanoparticle: a green$(+)$ arrow. An <u>increase</u> in the cooling rate decreases the nanoparticle temperature $T_{\mathrm{NP}}(t)$: a red$(-)$ arrow. An <u>increased</u> temperature has a very large influence on the atom loss (evaporation) rate: a thick green$(+)$ arrow. Finally, an <u>increased</u> atom loss rate strongly counteracts the growth of nanoparticles to a stable size $r^*$: a thick red$(-)$ arrow. In summary, with two red$(-)$ arrows in this chain, an increase in gas density should assist in the nucleation process.

We will in this section go through the numbered processes drawn in the diamonds in figure 2 one by one, and in Section 3 couple the whole system of processes together. The reader who first wants the broader picture can therefore go directly to Section 3.

### 2.1. Process 1, the connection between $p$, $n_{\mathrm{g}}$, and $T_{\mathrm{g}}$.

The process 1, symbolized by $n_{\mathrm{g}} = k_{\mathrm{B}} T_{\mathrm{g}}/p$ in figure 2, is the connection between the pressure, the temperature, and the gas density in zone 2. Due to the design of the experimental setup, the argon gas passes through the hollow cathode. Since the temperature of the hollow cathode surface is elevated, the gas would, between pulses, obtain a temperature $T_{\mathrm{g}}$ in the order of 1000 K [7] to 1500 K [8]. This temperature will be reduced by conduction when the gas has exited the hollow cathode. The energy dissipated when an amount of Ar gas, supplied through the hollow cathode, is cooled down from $T_{\mathrm{g}}$ to $T_{\mathrm{wall}}$, is given by:

$$E_{\mathrm{g}} = mc(T_{\mathrm{g}} - T_{\mathrm{wall}}) \tag{5}$$

where $m$ is the mass of the gas that has to be cooled down and $c$ is the specific heat capacity. The rate of heat conduction per unit area of the interface between zone 2 and zone 3 is given by Fourier´s law:

$$q = -k\nabla T \tag{6}$$

where $\nabla T \approx (T_{\mathrm{g}} - T_{\mathrm{wall}})/d$ is the temperature gradient across a thermal boundary of thickness $d$ between zones 2 and 3, and $k$ is the thermal conductivity, which is for a gas mixture of helium and argon is given by:

$$k_{\mathrm{mix}} = \sqrt{\frac{k_{\mathrm{B}}^3 T_{\mathrm{g}}}{\pi^3}}\left(\frac{1}{{d_{\mathrm{Ar}}}^2\sqrt{m_{\mathrm{Ar}}}(1+2.59\frac{X_{\mathrm{He}}}{X_{\mathrm{Ar}}})} + \frac{1}{{d_{\mathrm{He}}}^2\sqrt{m_{\mathrm{He}}}\left(1+0.7\frac{X_{\mathrm{Ar}}}{X_{\mathrm{He}}}\right)}\right) \tag{7}$$

where $X_{\mathrm{He}}$ is the mol fraction of helium and $X_{\mathrm{Ar}}$ is the mol fraction of argon and $k_B$is the Bolzmann constant [9].

The gas flow velocity is given by

$$v_{\mathrm{g}} = \frac{Q_{\mathrm{Ar}}\rho_{\mathrm{atm}}10^{-6}k_{\mathrm{B}}T_{\mathrm{g}}}{p\pi r_{\mathrm{gz}}^2 m_{\mathrm{Ar}}60} \tag{8}$$

where $r_{\mathrm{gz}}$ is the radius of the growth zone that the gas travels within, and $\rho_{\mathrm{atm}}$ is the density of the argon gas at atmospheric pressures [10]. An increased gas mass flow $Q_{\mathrm{Ar}}$ will increase the amount of gas atoms that has to be cooled down according to equation (5). From Eq. (6) it becomes evident that it takes time for the gas to cool down, and that this time depends on the mass density of the gas in zone 2, and on the heat conductivity in the thermal boundary to zone 3.

First let us consider the effect of changing the gas flow $Q_{\mathrm{Ar}}$ on the extent of zone 2, defined as the distance that the gas which exits the hollow cathode moves before it is cooled down. From Eq. (8) it is seen that the velocity of the gas increases proportionally to $Q_{\mathrm{Ar}}$. If the cooling rate according to Eq. (6) were constant (*i.e.,* independent of $Q_{\mathrm{Ar}}$), this would increase the extent of zone 2 proportionally to $Q_{\mathrm{Ar}}$. However, the thermal conductivity in zone 3 depends on the Ar/He gas mixture. The combined effect can be illustrated by a numerical example: if $Q_{\mathrm{Ar}}$ is increased from 10 to 20 sccm, the initial speed of the gas that has to be cooled down increases with a factor of two. If this were the only effect, the distance it moves before it has cooled down would increase by a about a factor of two. However, a higher $Q_{\mathrm{Ar}}$ also reduces the mole fraction of He in zone 3, and the thermal conductivity of Eq (6) decreases with about 20 %. This decreases the cooling rate and therefore further increases size of zone 2. As this numerical example shows, the extent of the hot zone 2 is expected increase a little more than proportionally with the gas flow $Q_{\mathrm{Ar}}$. This is in figure 2 drawn as a green(+) arrow from $Q_{\mathrm{Ar}}$ to $T_{\mathrm{g}}$.

The influence of $p$ on $T_{\mathrm{g}}$ is expected to be small. The argument goes briefly as follows. From Eq. (8) we see that a doubling of $p$ would reduce $v_{\mathrm{g}}$ with a factor of two. On the other hand, the density of the gas has doubled, which would increase the time it takes to cool it down, also by a factor of two. These two effects cancel. In the first approximation, we do not expect any effect from $p$ on the size of the hot zone 2, but a small such effect cannot be excluded. A dashed line with a question mark is therefore drawn in figure 2, from $p$ to $T_{\mathrm{g}}$, in order to indicate a possible influence.

Let us now look at the effect of adding a helium flow, $Q_{\mathrm{He}}$ to zone 3. A larger helium gas fraction gives a higher thermal conductivity in zone 3, and the gas is therefore cooled down faster in zone 2. This is probably the most important consequence of adding He because it has a large effect: if pure argon gas is substituted by pure helium in Eq. (7), there is a 775 % increase in the thermal conductivity.

Finally, we find that changing the chamber wall temperature will only have a small effect. It influences the gradient in Eq. (6). Assuming that the gas exiting the hollow cathode has a temperature of 1250 K, a decrease in the wall temperature from 425 K to 225 K only increases the cooling rate by 24 %.

In summary, we have now shown that $Q_{\mathrm{Ar}}$ is a dominating parameter for determining the extent of the hot zone 2, and thereby $T_{\mathrm{g}}$ close to the exit orifice of the hollow cathode.

## 2.2. Process 2, the expansion by diffusion of the ejected clouds of growth material

The process no 2 in figure 2 is diffusion of growth material. To estimate the densities of ions $n_{\mathrm{Ti}^+}$ and neutrals $n_{\mathrm{Ti}}$, the pulsed nature of the discharge has to be considered. First we have to estimate how many ions and neutral atoms that are ejected out from the hollow cathode. This is done by the following relation for ions,

$$N_{\mathrm{Ti}^+\mathrm{Pulse}} = \frac{(\int I_{\mathrm{Pulse}} dt) Y_{\mathrm{sput}} f_{\mathrm{Ti}^+} f_{\mathrm{ext,Ti}^+}}{e} \approx 8.22 \times 10^{13} \qquad (9)$$

and an analogous relation for Ti atoms. Here $(\int I_{\mathrm{Pulse}} dt) \approx 2.7 \times 10^{-4}$ is the integrated current of one pulse, $e$ is the unit for electric charge, and $Y_{\mathrm{sput}} \approx 0.35$ is the sputter yield [11]. $f_{\mathrm{Ti}^+}$ is the fraction of the sputtered material that becomes ionized (which is much higher in this high power pulsed discharge than in usual hollow cathode discharges), and $f_{\mathrm{ext,Ti}^+}$ is the fraction of these ions that get extracted from the hollow cathode. For these fractions we have no measurements, and the only theoretical estimates are from a model of a similar discharge, but with a Cu cathode, by Hasan *et al* [12]: $f_{\mathrm{ext,Ti}^+} \approx 17$ % for ions, $f_{\mathrm{ext,Ti}} \approx 3$ % for neutrals, and $f_{\mathrm{Ti}^+} \approx 80$ %, giving $f_{\mathrm{Ti}} = (1 - f_{\mathrm{Ti}^+}) \approx 20$ %. The last step in Eq. (9) is a very approximate estimate for our discharge, using these values. By assuming that each pulse ejects the quantities $N_{\mathrm{Ti}^+,\mathrm{pulse}}$ of ions and $N_{\mathrm{Ti,pulse}}$ atoms that follow the gas flow out of the hollow cathode while they expand by diffusion as a spherical cloud. The radial distribution is then [13] a Gaussian with the scale radius $R = \sqrt{2D\Delta t}$, at which there is an e-fold decrease in density from the central value. $D$ is the diffusion coefficient, and $\Delta t$ is the time elapsed after the moment of initiation. The time dependent space-average density in the cloud is approximated as exemplified here for the neutrals, through dividing the total number of atoms with a characteristic volume

$$n(t) = \frac{3N_{\mathrm{Ti,pulse}}}{4\pi(2Dt+r_{\mathrm{hc}}{}^2)^{3/2}}, \tag{10}$$

where $t$ is the time after the cloud center exiting the hollow cathode, and $r_{\mathrm{hc}}$ is the radius of the hollow cathode which is taken to be the radius of the expanding cloud as it exits the orifice. The diffusion coefficient of neutrals $D$ is from classical diffusion. Ions diffuse at a higher rate given by the ambipolar diffusion coefficient $D_{\mathrm{a}}$. Close to the exit of the hollow cathode the gas is approximated to be pure argon (with a low degree of ionization), and the diffusion coefficient is given by [14]:

$$D = \frac{v_{\mathrm{th}} l_{\mathrm{coll}}}{3} = \frac{2}{3}\sqrt{\frac{k_{\mathrm{B}}^3}{\pi^3}}\sqrt{\frac{1}{2m_{\mathrm{Ar}}} + \frac{1}{2m_{\mathrm{Ti}}}}\frac{4T_g^{3/2}}{p(d_{\mathrm{Ar}}+d_{\mathrm{Ti}})^2} \tag{11}$$

and the ambipolar diffusion coefficient by:

$$D_{\mathrm{a}} = \frac{v_{\mathrm{i,th}} l_{\mathrm{coll}}}{3}\sqrt{\frac{T_{\mathrm{e}}+T_{\mathrm{i}}}{T_{\mathrm{i}}}} = \sqrt{\frac{8k_{\mathrm{B}}T_{\mathrm{i}}}{\pi m_{\mathrm{Ti}}}}\frac{k_{\mathrm{B}}T_{\mathrm{g}}}{3p\sigma_{\mathrm{Ti^+}}}\sqrt{\frac{T_{\mathrm{e}}+T_{\mathrm{i}}}{T_{\mathrm{i}}}} \tag{12}$$

where $l_{\mathrm{coll}} = \frac{1}{n_{\mathrm{g}}\sigma_{\mathrm{Ti^+}}} = \frac{k_{\mathrm{B}}T_{\mathrm{g}}}{p\sigma_{\mathrm{Ti^+}}}$ is the collision mean free path, $v_{\mathrm{th}}$ and $v_{\mathrm{i,th}}$ are the thermal velocities of neutrals and ions, $m_{\mathrm{Ar}}$ is the mass of an argon atom, $m_{\mathrm{Ti}}$ is the mass of a titanium atom, $d_{\mathrm{Ar}}$ is the collision diameter of an argon atom, and $d_{\mathrm{Ti}}$ is the collision diameter of a titanium atom which was estimated to be $\sim 3 \times 10^{-10}$ [m]. For the ambipolar diffusion $T_{\mathrm{e}}$ is the electron temperature, $T_{\mathrm{i}}$ is the ion temperature, and $\sigma_{\mathrm{Ti^+}}$ is the elastic collision cross section between a titanium ion and an argon atom, for which the same value as in [5] was used.

Process 2 in figure 2, diffusion of growth material, is a key process for the reason that it determines how the densities $n_{\mathrm{Ti}}$ and $n_{\mathrm{Ti^+}}$ evolve in time in the ejected clouds.

### 2.3. Process 3, cooling collisions with process gas

The nanoparticle temperature is determined by $T_{\mathrm{g}}$ plus a heating contribution from exothermic reactions on the nanoparticle surface [15]. Figure 3 (which will be fully described in section 2.7) illustrates what we call single-heating events, in which there are momentary temperature increases followed by time decreases to an equilibrium temperature close to $T_{\mathrm{g}}$.

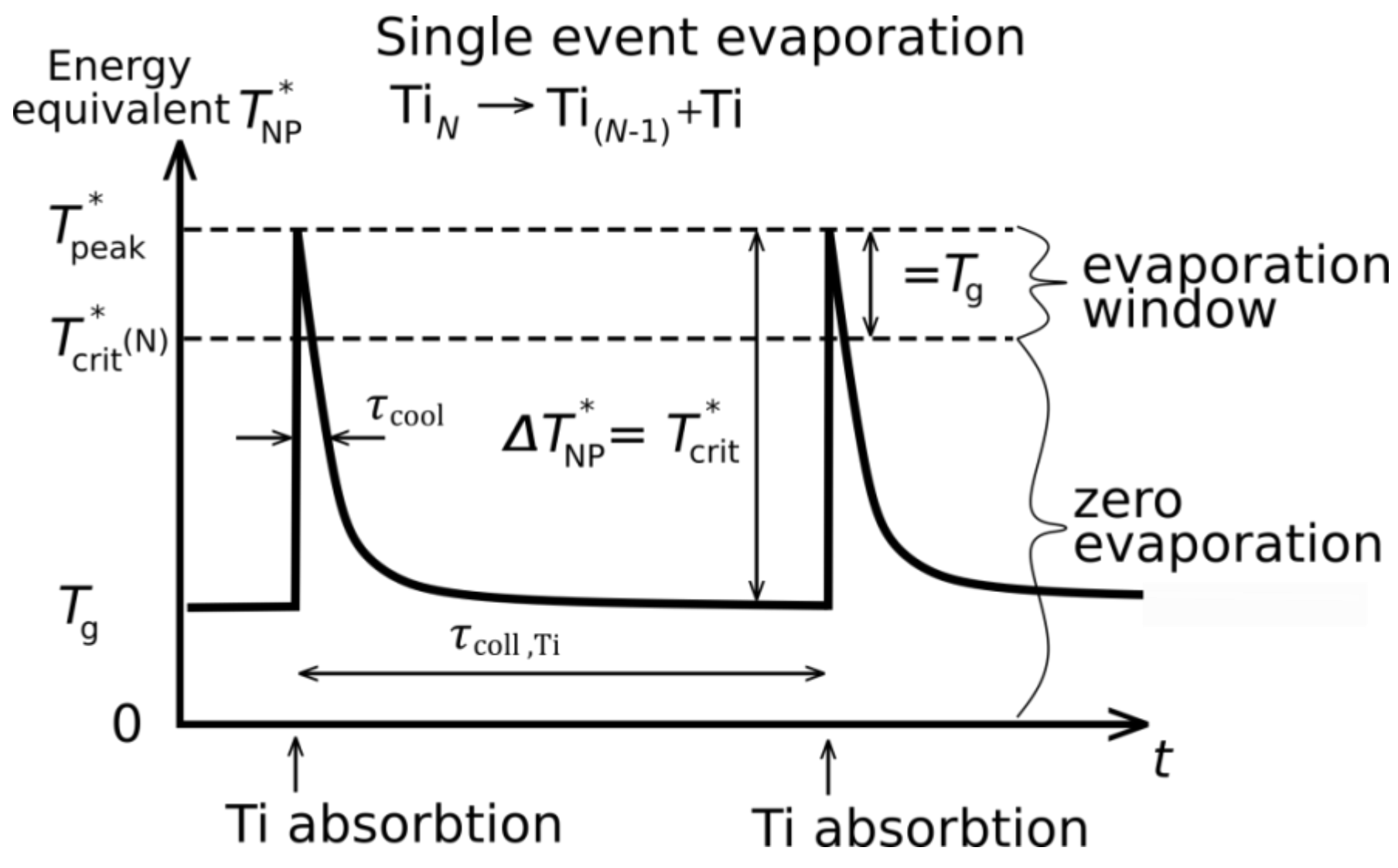

*Figure 3 A schematic illustration of how the energy-equivalent temperature $T_{NP}^*$ $(a\ concept\ to\ be\ discussed\ in\ section\ 2.7)$ of a small ($N \leq 10$) nanoparticle varies after absorption of a Ti atom. Evaporation is only possible in a small evaporation window which has a width that can be approximated by the gas temperature $T_g$.*

A simple estimate can be made to determine if this is the situation during the nucleation phase in our experiments. If we approximate that a nanoparticle which contains $N$ atoms will be significantly cooled when it has collided with $N$ process gas atoms, we get the cooling time as $\tau_{\text{cool}} \approx N\tau_{\text{coll,gas}}$. The time it takes before it collides with and absorbs another Ti atom is $\tau_{\text{coll,Ti}}$. When $\frac{\tau_{\text{cool}}}{\tau_{\text{coll,Ti}}} \ll 1$ , we have single-heating evens. The collision times are to the first approximation inversely proportional to the densities of the colliding species, giving $\frac{\tau_{\text{coll,gas}}}{\tau_{\text{coll,Ti}}} \approx \frac{n_{\text{Ti}}}{n_{\text{gas}}}$. From the relations above, the condition for single-heating events becomes

$$\frac{\tau_{\text{cool}}}{\tau_{\text{coll,Ti}}} \approx N\frac{n_{\text{Ti}}}{n_{\text{g}}} \ll 1 \tag{13}$$

Typical values in our experiment are a process gas density (argon plus helium) of $n_{\text{g}} \approx 10^{22}$ m$^{-3}$ and $n_{\text{Ti}} \approx 10^{19}$ m$^{-3}$ and thus the condition of Eq. (13) is satisfied for small nanoparticles. Particles smaller than 1 nm are usually referred to as clusters, but for simplicity reasons we will continue calling them nanoparticles.

The cooling rate in a single-heating event is determined by heat transfer between the gas and the nanoparticle. In the free molecular regime (where the mean free path is $\gg r_{\text{NP}}$), as viewed from the perspective of the nanoparticle, the formula for heat transfer from a nanoparticle to the surrounding gas is given by

$$q = \alpha\pi r_{\text{NP}}^2 p\sqrt{\frac{2k_{\text{B}}T_{\text{g}}}{\pi m_{\text{g}}}}\left(\frac{\kappa+1}{\kappa-1}\right)\left(\frac{T_{\text{NP}}^*}{T_{\text{g}}} - 1\right) \tag{14}$$

where $r_{\mathrm{NP}}$ is the radius of the nanoparticle, $m_{\mathrm{g}}$ is the mass of the gas atom, $T_{\mathrm{NP}}^{*}$ is the temperature of the nanoparticle, and $\kappa$ is the specific heat ratio [16]. The constant $\alpha$ is the thermal accommodation coefficient, which depends on which type of gas atom that collides with the particle. For collisions with a stainless steel surface, values of $\alpha = 0.866$ for argon and $\alpha = 0.360$ for helium have been measured [17]. This difference in $\alpha$ counteracts the higher speed of the He atoms, with the result that if the argon gas is substituted by helium, there is only a 31 % increase in the cooling rate of the nanoparticle. This difference is so small that it is unimportant to consider the He/Ar fraction of the gas for the cooling effect, only the total number density $n_{\mathrm{g}}$ is needed. From Eq. (14) follows that the nanoparticle's cooling rate $q$ will be proportional to $p$. The temperature effect is that, during the time the nanoparticle is hot in the sense that $T_{\mathrm{NP}}^{*}/T_{\mathrm{g}} \gg 1$, the first approximation is that $q \propto \sqrt{T_{\mathrm{g}}}$ at constant $p$. This is because there is only a small effect due to the last parenthesis in Eq. (14) which accounts for the difference between $T_{\mathrm{g}}$ and $T_{\mathrm{NP}}^{*}$. For a numerical example: if $T_{\mathrm{NP}}^{*}$ is 1500 K a change in $T_{\mathrm{g}}$ by a factor of two, from 300 to 600 K, changes the rate of cooling by collisions by only 53 %. This influence is therefore indicated by a dashed line in figure 2.

In summary regarding process 3 in the flow chart of Figure 2: the gas-collision cooling rate is increased by higher $n_{\mathrm{g}}$ (a green(+) arrow to process 3) and, somewhat weaker, decreased by increased $T_{\mathrm{g}}$ (a dashed red(-) arrow to process 3). The main effect of the cooling collisions on the nucleation process is that faster cooling, in the time-dependent $T_{\mathrm{NP}}^{*}(t)$, reduces the time duration of the evaporation window (a red(-) arrow from process 3).

### 2.4. Process 4, the value of the "baseline" temperature $T_{\mathrm{NP}}^{*} = T_{\mathrm{g}}$

This process is given a separate place in the flow chart of figure 2 for the reason that the baseline temperature of the nanoparticles, before a single-heating event, it is the most important parameter for the growth probability from dimers to stable size $r*$. The cooling process itself is very simple. The collisional cooling gives the nanoparticles an equilibrium temperature $T_{\mathrm{NP}}^{*} = T_{\mathrm{g}}$ after a few cooling times $\tau_{\mathrm{cool}}$ as shown in figure 3. An approximate formula for the cooling time is

$$\tau_{\mathrm{cool}} \approx \frac{N\tau_{\mathrm{coll,gas}}}{\alpha}\,. \tag{15}$$

With parameters from zone 2 for our experimental conditions, and assuming $N < 10$ during the nucleation phase, we obtain $\tau_{\mathrm{cool}}$ to be typically less than 100 ns. The approach to $T_{\mathrm{NP}}^{*} = T_{\mathrm{g}}$ is marked with a green(+) arrow from process 4 in the flow chart of figure 2.

### 2.5. Dimer formation

The formation of dimers is a necessary first step for the growth of nanoparticles, and often considered as the bottleneck since it is not possible to form a dimer in a two-body collision between two atoms. The reason is that the dimers internal vibrational energy must be lower than the binding energy for it to be bound. Since it is precisely the binding energy that is released when the dimer is formed, some energy has to be removed, and for this a third body is needed. If the initial kinetic energy (due to thermal motion) in the rest frame of the dimer-forming particles is $E_{\mathrm{th}}$, then this third body must carry away more than the energy $E_{\mathrm{th}}$.

### 2.5.1. Two-body dimer formation in an oxygen-rich environment

The two-body dimer formation is not included in the flow chart of figure 2 which refers to an oxygen-starved environment. Dimer formation by two-body collisions is possible if the titanium atom or ion collides with a molecule which splits apart. The separated atom or molecule can then carry away enough kinetic energy to make the remaining dimer stable. In our case, the most likely such reaction from residual gases in the vacuum system is

$$\mathrm{Ti}^{+} + \mathrm{H_2O} \rightarrow \mathrm{TiO}^{+} + \mathrm{H_2}. \qquad (16)$$

The expression for this two-body collision rate is taken from [18] as $R_{\mathrm{TiO}^{+}} = \sigma_{\mathrm{Ti}^{+}\mathrm{H_2O}} v_{\mathrm{rel}} n_{\mathrm{Ti}^{+}} n_{\mathrm{H_2O}}$, from which we get the collision frequency of a $\mathrm{Ti}^{+}$ ion by dividing with $n_{\mathrm{Ti}^{+}}$,

$$f_{\mathrm{Ti}^{+}\mathrm{H_2O}} = \sigma_{\mathrm{Ti}^{+}\mathrm{H_2O}} v_{\mathrm{rel}} n_{\mathrm{H_2O}}. \qquad (17)$$

Here, $n_{\mathrm{H_2O}}$ is the density of water vapor in the vacuum system, $\sigma_{\mathrm{Ti}^{+}\mathrm{H_2O}}$ is the cross section for collisions between a titanium ion and a water molecule, and $v_{\mathrm{rel}}$ is the relative collision velocity between a titanium atom and a water molecule given by

$$v_{\mathrm{rel}} = \left(\frac{8 k_{\mathrm{B}} T_{\mathrm{g}}}{\pi \mu}\right)^{1/2} \qquad (18)$$

where μ is the reduced mass of the two colliding species. For a numerical example we use the parameters in the experiments in the high vacuum system used in [5], at the oxygen limit that is marked (1) in figure 1 (c). We are interested in the fraction of the $\mathrm{Ti}^{+}$ ions that form TiO dimers during the time $\tau$ they are in zone 2, after leaving the hollow cathode. We take the ejection speed of the growth material to be $v_z \approx$ 100 m/s from Hasan *et al* [12], and consider a distance of 1 cm. This gives $\tau \approx \frac{0.01}{100}$ =100 µs. For an individual $\mathrm{Ti}^{+}$ ion, the probability of forming a dimer at the base pressure ($n_{\mathrm{H_2O}}$ = $3.8 \cdot 10^{15}\ \mathrm{m^{-3}}$), within 1 cm, then becomes $\tau \times f_{\mathrm{Ti}^{+}\mathrm{H_2O}}$ = 100 $\times 10^{-6} \times 5.9 \times 10^{-19} \times 1422 \times 3.8 \times 10^{15} = 3 \times 10^{-4}$. This means that about 0.03 % of the $\mathrm{Ti}^{+}$ ions will form TiO dimers within 1 cm from the hollow cathode. With $N_{\mathrm{Ti^{+}Pulse}}$ from Eq. (9), the number of TiO dimers formed in each pulse becomes $N_{\mathrm{TiO\ Pulse}} \approx 3 \times 10^{-4} \times 8.22 \times 10^{13} \approx 10^{10}$. Although this number is very uncertain, due to a large error factor in $n_{\mathrm{H_2O}}$ as discussed in Appendix 4, it is sufficiently accurate to exclude dimer formation as the bottleneck in nanoparticle formation. The argument goes as follows. If all the $10^{10}$ dimers formed were to grow to our typical 30 nm nanoparticles, then ten pulses would be enough to form a monolayer of nanoparticles on a substrate of 1 cm$^2$. With the used frequency 1500 Hz, 10 minutes of operation corresponds $\approx 10^5$ monolayers which is several orders of magnitude more than what was found. The bottleneck that causes the disappearance of nanoparticles at the oxygen limit has to be in a later stage of the nanoparticle growth and collection.

### 2.5.2. Process 5, three-body dimer formation

In our discharges there is a large degree of ionization of the growth material. In the oxygen starved environment the main candidate for three-body dimer formation is therefore

$$\mathrm{Ti}^+ + \mathrm{Ti} + \mathrm{Ar} \rightarrow \mathrm{Ti}_2^+ + \mathrm{Ar} \tag{19}$$

where a titanium ion and a titanium atom collide, at the same time as they collide with an argon atom. The titanium ion and atom can then bind together if the argon atom carries away enough excess kinetic energy. The expression for the rate for this three-body collision is taken from Smirnov [19], and adapted to fit the highly ionized plasma in the current experimental setup (for details, see Appendix 1):

$$R_{\mathrm{Ti}_2{}^+} = n_{\mathrm{Ti}} n_{\mathrm{Ti}^+} n_{\mathrm{Ar}} v_{\mathrm{relAr}} b^3 \sigma_{\mathrm{Ti}^+\mathrm{Ar}}. \tag{20}$$

Here $n_{\mathrm{Ti}}$ is the density of titanium, $n_{\mathrm{Ti}^+}$ is the density of titanium ions, $n_{\mathrm{Ar}}$ is the density of argon atoms, $v_{\mathrm{relAr}}$ is the relative collision velocity between an argon atom and a titanium atom, $\sigma_{\mathrm{Ti}^+\mathrm{Ar}}$ is the cross section for collisions between argon neutrals and the titanium ion, and $b$ is the critical radius for interaction between a titanium ion and a titanium neutral (see Appendix 1). Combining equations (9), (10), (A3), (A4), and (20) we get the rate of $\mathrm{Ti}_2$ dimer formation by three-body collisions as

$$R_{\mathrm{Ti}_2{}^+} = \left[\frac{3N_{\mathrm{Ti\,pulse}}}{4\pi(2Dt+r_{\mathrm{hc}}{}^2)^{3/2}}\right]\left[\frac{3N_{\mathrm{Ti}^+\mathrm{pulse}}}{4\pi(2D_{\mathrm{a}}t+r_{\mathrm{hc}}{}^2)^{3/2}}\right]\left[\frac{p}{k_{\mathrm{B}}T_{\mathrm{g}}}\right]\left[\left(\frac{8k_{\mathrm{B}}T_{\mathrm{g}}}{\pi\mu}\right)^{\frac{1}{2}}\left(\frac{R_0{}^4 6\epsilon(1-\gamma)}{k_{\mathrm{B}}T_{\mathrm{g}}}\right)^{3/4}\pi\sqrt{\frac{\alpha_{\mathrm{Ar}}q^2}{\epsilon_0 8k_{\mathrm{B}}T_{\mathrm{g}}}}\right] \tag{21}$$

For easy reference to the process flow chart in figure 2, the right-hand side is written as the product of four square brackets. From the left to the right, these four brackets correspond to the four arrows (from top to down) drawn to process 5, three-body collisions. In the same order, the arrows (and the brackets) represent the factors $n_{\mathrm{Ti}}$, $n_{\mathrm{Ti}^+}$, $n_{\mathrm{Ar}}$, and $v_{\mathrm{relAr}} b^3 \sigma_{\mathrm{Ti}^+\mathrm{Ar}}$ in the reaction of Eq. (21).

Using Eq (21) we now can connect the dimer formation rate with the pressure $p$ and $T_{\mathrm{g}}$ in zone 2. From combining Eq:s (11), (12) and (21), we see that the amount of dimers that can be created per second is $\propto p^4$ and $\propto T_g^{-5.5}$ (we estimate[1] that these relations apply when the cloud has expanded to have approximately twice the size of the hollow cathode orifice, which in Eq. (21) corresponds to $2Dt > r_{hc}^2$). It is thus crucial to reduce $T_g$ and increase $p$ to create an environment which promotes three-body dimer formation. It is worth noting that the major part of the temperature dependence is indirect, in the sense that it comes from the first two square brackets in Eq (21) which reflect the density of growth material. The physical process can easily be followed in figure 2: increasing $T_g$ at constant $p$ decreases $n_g$, which increases the diffusion rates (both ordinary and ambipolar). Faster diffusion, in turn, decreases $n_{Ti^+}$ and $n_{Ti}$.

### 2.6. Process 6, atomistic growth

The process number 6 in figure 2, the growth from dimers to stable size $r^*$ is through adding single Ti atoms, also called atomistic growth.

$$\mathrm{Ti}_N + \mathrm{Ti} \rightarrow \mathrm{Ti}_{(N+1)} \qquad (22)$$

Coagulation of nanoparticles is often proposed to be an important growth mechanism but can be neglected in our type of pulsed experiment. The reason is the short effective time available. For a numerical example we take the value from section 2.5.1, where a 0.03 % fraction of the growth material was estimated to form dimers within 1 cm from the hollow cathode orifice. Even if each of these dimers would initiate the growth of a nanoparticle, the nanoparticle density will be about a factor 360 below the density of Ti atoms and 8220 below the density of Ti ions. A given individual nanoparticle is therefore more likely to collide with single atoms or ions, than to coagulate with earlier formed nanoparticles. Also growth by addition of $Ti^+$ ions can very likely be neglected, for reasons detailed in Appendix 2. In short, for small sub critical nanoparticles this process heats the nanoparticles so much that, when a $Ti^+$ ion is added, then a Ti atom is likely to be lost very soon by evaporation.

This leaves nucleation through atomistic growth according to Eq. (22). To analyze this process we need the binding energy of the last added atom. Hybrid density functional theory (DFT) ab initio quantum-chemical computations were therefore carried out (for details, see Appendix 3) in order to obtain binding energies in small titanium and titanium oxide nanoparticles of atoms, which compose our models for the nanoparticles. The results are given in Table I for neutral nanoparticles, and in Table II for charged nanoparticles.

If we first look at the binding energy of neutral nanoparticles only containing titanium, we see that the energy of $Ti_2$ is only 0.83 eV. This means that they are very susceptible to splitting apart at elevated temperatures. The average binding energy when adding a titanium atom to a growing titanium nanoparticle, larger than $Ti_2$, is 2.48 ± 0.66 eV for all sizes investigated, up to $Ti_{16}$. There is one statistical outlier, $Ti_{13}$ which has a significantly higher binding energy of 3.62 eV. This nanoparticle in its ionized form has also been observed to be a magic number [20] and its high binding energy is probably due to its icosahedron shape.

[1] The approximation of a spherical and expanding cloud, represented by Eq. (10), is the basis for the first two factors (the square-bracket parentheses) in Eq. (21). This is not a good approximation close to the hollow cathode orifice, but becomes better further away from it.

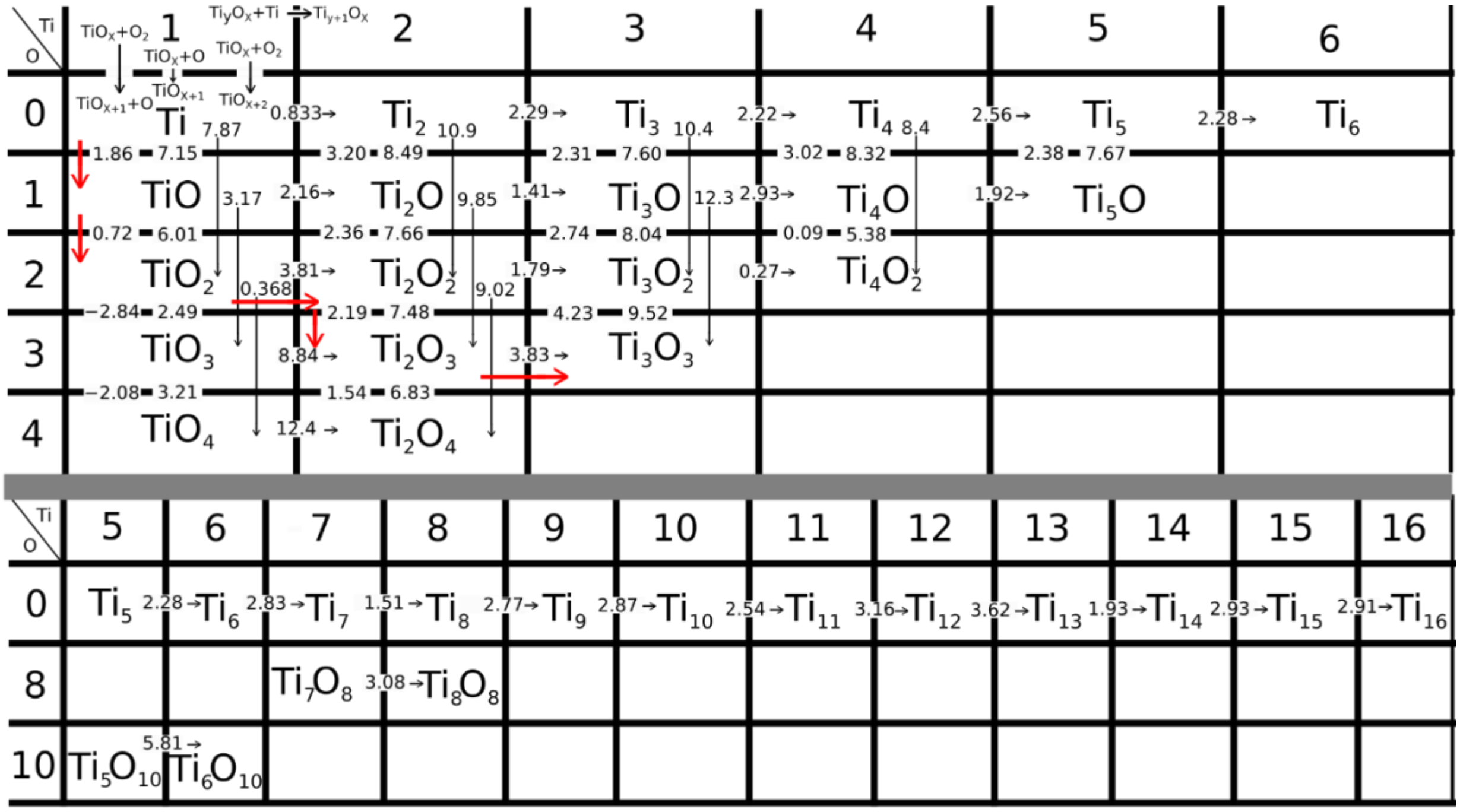

*Table 1 Binding energy when adding a titanium atom (arrows to the right). Binding energy when adding an oxygen atom (middle of lower boundary to each cell). Energy released when adding an oxygen molecule (2 cells down, long arrow). Net energy released when adding an oxygen molecule followed by evaporation of one oxygen atom (left lower corner of each cell). The red arrows denote the path of highest binding energy in an oxygen rich environment. It can be seen that the energy released when adding a titanium atom is generally lower than when an oxygen atom is added.*

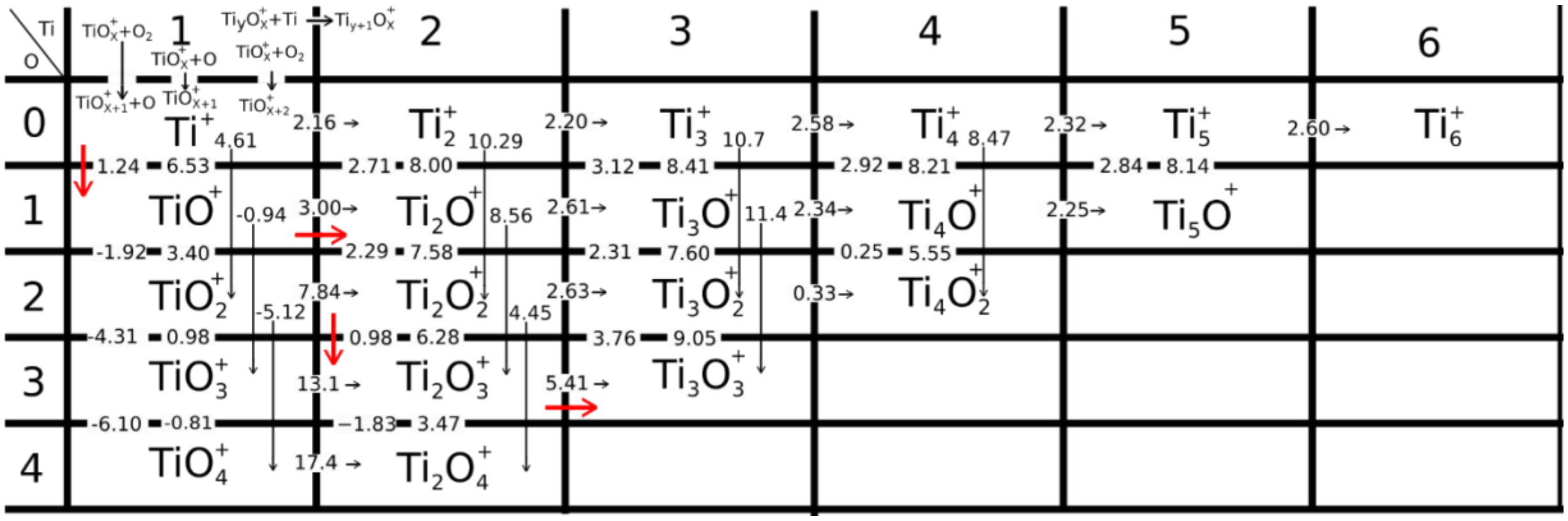

*Table 2 Binding energies of ionized nanoparticles for added Ti atoms, O atoms, and $O_2$ molecules, denoted as in Table 1. The red arrows denote the most stable route of growth.*

It can also be seen that the binding energies of subsequent Ti atoms, of a growing nanoparticle that started as a TiO dimer, is not significantly higher than for one that started as a $Ti_2$ dimer.

This shows that one single oxygen atom does not help the growth beyond the first (dimer formation) growth stage. For oxygen to significantly help the nanoparticles reach $r*$, there has to be an abundance of it in order for them to grow along the red arrows.

If we now focus on the ionized nanoparticles in table 2, we see that the same general trends as for neutral particles hold. However, the binding energy of a $Ti_2^+$ dimer (2.16 eV) is significantly higher than that of $Ti_2$ (0.833 eV).

### 2.7. Process 7, titanium atom loss

The process numbered 7 in figure 2 is the evaporation of single titanium atoms from a pure titanium nanoparticle. We will here first give the established thermodynamic description of this process, and then show that this description needs to be significantly modified for the very small nanoparticles during growth from dimers to $r*$.

Mangolini *et al* [15] explained the heating behavior of small nanoparticles as a result of exothermic reactions on the nanoparticle surface. They found that $T_{\mathrm{NP}}^*$ can exceed $T_{\mathrm{g}}$ by several hundreds of Kelvins during short periods of time and then, with the cooling by the gas, get back to temperatures as low as the gas. This is the same type of time evolution as shown above in figure 3. If $T_{\mathrm{g}}$ were to be 300 K higher, the peak heights after an exothermic reaction would also be 300 K higher. This has a surprisingly large effect on the evaporation rate. For a qualitative demonstration, we make a thought experiment with hypothetical nanoparticles that have the same properties as bulk material. The vapor pressure of bulk material can be approximated by the Antonine equation:

$$p_{vap} = 10^{\mathrm{A}-\frac{\mathrm{B}}{T_{\mathrm{NP}}}} \tag{23}$$

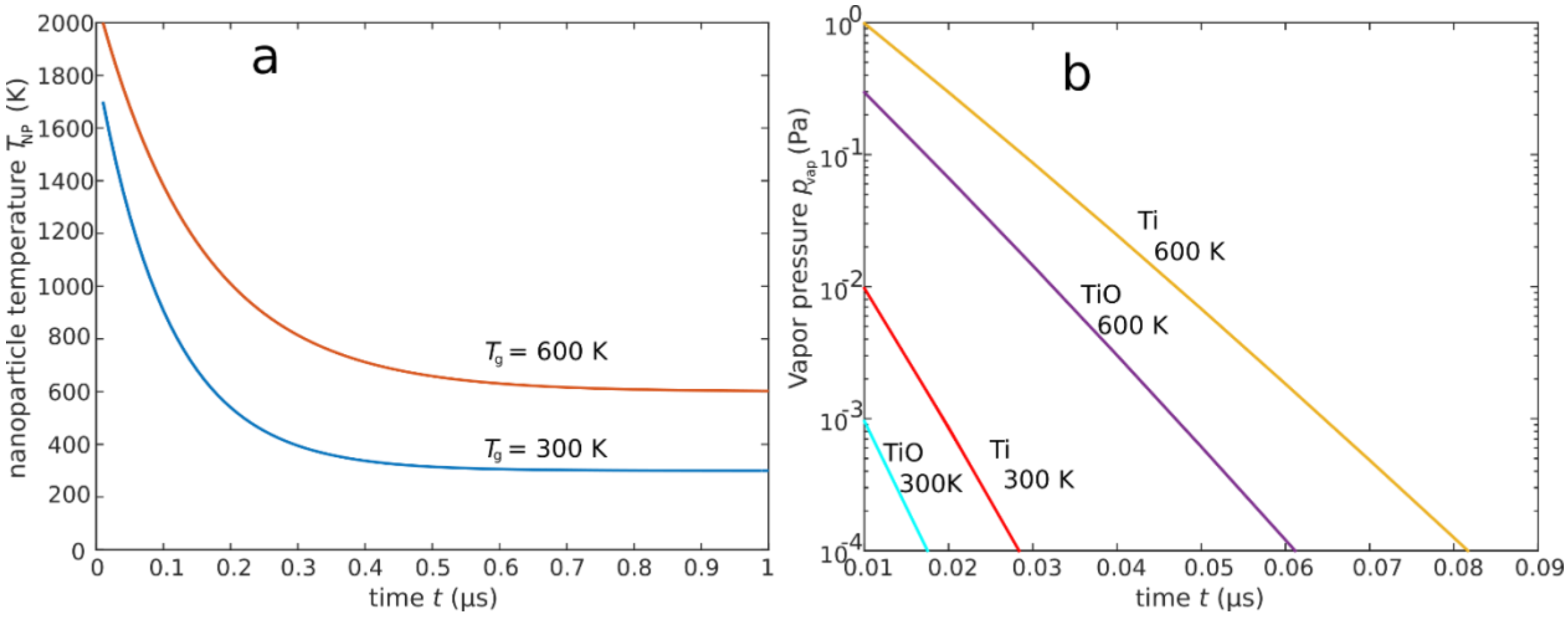

*Figure 4 The time-dependent evaporation of hypothetical titanium and titanium oxide nanoparticles with the same vapor pressure as bulk material. (a) The temperature $T_{NP}(t)$ after an exothermic reaction which has heated a* $Ti_{15}$ *nanoparticle by 1400 K. The nanoparticle*

*temperature at two different gas temperatures are shown. The temperature is at all times higher when the gas temperature is higher. (b) The vapor pressures a function of time for* Ti *and* TiO*, with $T_{NP}(t)$ taken from panel 4(a).*

where A and B are element specific constants. When a nanoparticle is cooled according to Eq. (14), the evaporation rate will depend on the continuously decreasing temperature. In figure 4 (a) the cooling of a nanoparticle with the size and thermal mass of 15 atoms is plotted with the approximation that the nanoparticle has the same evaporation constants A and B as bulk material. The pressure is chosen to be 300 Pa in a pure helium atmosphere. This figure illustrates a typical cooling behavior after an exothermic reaction which heats the nanoparticle by 1400 K. If $T_{\mathrm{g}}$ is 300 K, this event will elevate $T_{\mathrm{NP}}^{*}$ to 1700 K. If instead $T_{\mathrm{g}}$ is 600 K, the peak $T_{\mathrm{NP}}^{*}$ will be 2000 K. The difference is only 15% but it has a profound influence on the vapor pressure which is shown in Figure 4(b). In this example, where $T_{\mathrm{g}}$ is increased by a factor of two, the peak $T_{\mathrm{NP}}^{*}$ is increased by only 15%, while the peak vapor pressure increases 100 times, from 0.01 Pa to 1 Pa. From the figure, it can also be seen that the titanium oxide nanoparticles have a lower vapor pressure, and can thus withstand heat better without evaporating.

Let us now turn to the modification of this classical evaporation model. To this purpose we rewrite Eq (23) in a mathematically equivalent form obtained from statistical decay theory, which we take from Borggreen *et al* [21]. The rate for the evaporation from a $\mathrm{Ti}_N$ nanoparticle at a temperature $T_{\mathrm{NP}}$ is then assumed to be dependent on the internal energy only, which gives

$$k(N, T_{\mathrm{NP}}) = A(N) \exp\left(-\frac{E_{\mathrm{evap}}(N)}{k_{\mathrm{B}} T_{\mathrm{NP}}}\right) \tag{24}$$

in units of evaporated particles per second. $A(N)$ is a constant that depends on the size of the nanoparticle, and the evaporation energy $E_{\mathrm{evap}}$ is the binding energy of the weakest bound $N$th atom. $E_{\mathrm{evap}}$ is generally smaller for smaller nanoparticles than for bulk material, leading to the common opinion [22] that the vapor pressure should be higher for smaller nanoparticles. We will here argue against this assumption. We will draw conclusions that seems counter-intuitive to basic thermodynamics. In order to pinpoint the problem, we will therefore first present an apparent paradox.

The apparent paradox concerns the vacuum pressure of a $\mathrm{Ti}_4$ nanoparticle. From the exponential factor in Eq. (24) follows that it should be a function of the ratio $\frac{E_{\mathrm{evap}}}{T_{\mathrm{NP}}}$. For bulk titanium at the boiling temperature 3560 K we get the ratio $\frac{E_{\mathrm{evap}}}{T_{\mathrm{bulk}}} = \frac{4.4}{3560} = 0.0012$. At this ratio of $\frac{E_{\mathrm{evap}}}{T_{\mathrm{bulk}}}$ titanium therefore has, by definition, a vapor pressure of 1 atm. We now consider a $\mathrm{Ti}_4$ nanoparticle at a temperature we choose to be $T_{\mathrm{NP}}$ = 1780 K. Table I gives $E_{\mathrm{evap}} = 2.22$ eV and therefore $\frac{E_{\mathrm{evap}}}{T_{\mathrm{NP}}} = \frac{2.22}{1780} = 0.0012$. This is the same value as bulk titanium had at the boiling temperature. Consequently, one would, based on the thermodynamic equation (24), expect a $\mathrm{Ti}_4$ nanoparticle at $T_{\mathrm{NP}} = 1780$ K to have a vapor pressure of one atmosphere.

Now let us consider a very specific way to construct a $\mathrm{Ti}_4$ nanoparticle with a temperature in this range. We start with a $\mathrm{Ti}_3$ nanoparticle with zero internal energy ($T_{\mathrm{NP}}^* = 0$). Then we add one titanium atom and let the energy $E_{\mathrm{evap}} = 2.22$ eV be converted to heat. Finally, we let one gas atom collide with the nanoparticle and cool it a little, removing 0.02 eV. Now, this nanoparticle has an internal thermal energy of $E_{\mathrm{th}} = 2.20$ eV, and it would require 2.22 eV to evaporate one titanium atom. Evaporation is energetically impossible. The vapor pressure is therefore exactly zero. In order to get a ballpark estimate of the temperature of this nanoparticle, we can approximate that it has the same specific heat capacity as bulk titanium, 544.3 [J/(kg K)]. This gives the internal thermal energy $E_{\mathrm{th}} = 3.16 k_{\mathrm{B}}$ per atom and K. With four atoms in the nanoparticle, the temperature becomes approximately

$$T_{\mathrm{NP}}^* \approx \frac{2.20e}{4 \times 3.16 k_{\mathrm{B}}} = 2021\ \mathrm{K} \tag{25}$$

The paradox is this: from the thermodynamic equation (24), we estimated that a $\mathrm{Ti}_4$ nanoparticle at 1780 K should have 1 atm vacuum pressure. On the other hand, directly from energy conservation, we can prove that a $\mathrm{Ti}_4$ nanoparticle with a temperature about 2021 K has exactly zero vapor pressure. Which is the correct estimate, and why is there a difference?

The solution to this apparent paradox lies in that the two cases implicitly use two different definitions of the concept "nanoparticle temperature", which we have highlighted above by the use of two different variables $T_{\mathrm{NP}}$ and $T_{\mathrm{NP}}^*$. The thermodynamic temperature $T_{\mathrm{NP}}$ is a quantity that is valid for an ensemble of nanoparticles which is in equilibrium with a heat bath. Such an ensemble has a spread in internal energy, including a high-energy thermal tail. By contrast, nanoparticles that all have obtained the same amount of internal energy, in this case 2.22 eV by absorbing a titanium atom, all have the same internal energy. For such a case we herein use the energy-equivalent temperature $T_{\mathrm{NP}}^*$, defined as the temperature of a heat bath which would give them the correct average internal energy. The difference in vapor pressure arises from the fact that only the thermodynamic ensemble has a high-energy tail, and it is this tail which gives it a higher vapor pressure.

For small enough nanoparticles, the pre-exponential factor in Eq. (24) therefore depends on whether $T_{\mathrm{NP}}$ or the energy-equivalent $T_{\mathrm{NP}}^*$ is relevant in the studied case. For single-event evaporation events such as we study here, $T_{\mathrm{NP}}^*$ must be used. In this case the pre-exponential factor becomes temperature dependent, and the substitution $A(N) \rightarrow A(N, T_{\mathrm{NP}}^*)$ must be made in Eq. (24). The paradox above illustrates that there then exists a critical temperature such that $A(N, T_{\mathrm{NP}}^*) = 0$ for $T_{\mathrm{NP}}^* \leq T_{\mathrm{crit}}^*$ because the internal thermal energy is insufficient for vaporization of even one atom. $T_{\mathrm{crit}}^*$ is easily obtained from the condition that the thermal energy equals the energy $E_{\mathrm{evap}}$ needed for removal of the weakest bound titanium atom, giving

$$T_{\mathrm{crit}}^* = \frac{E_{\mathrm{evap}}}{\vartheta k_{\mathrm{B}} N} \tag{26}$$

where $\vartheta$ is the specific heat capacity per atom.

Let us now consider how the existence of a critical temperature influences single-event evaporation such as show in figure 3. The nanoparticle is here assumed to have the gas temperature $T^*_{\mathrm{NP}} = T_{\mathrm{g}}$ before the event. When a titanium atom is absorbed, the binding energy $E_{\mathrm{evap}}$ is converted into heat. From the definition of $T^*_{\mathrm{crit}}$ follows that this gives the nanoparticle an increase $\Delta T^*_{\mathrm{NP}} = T^*_{\mathrm{crit}}$ to the new temperature[2] $T^*_{\mathrm{peak}} = T_{\mathrm{g}} + T^*_{\mathrm{crit}}$. The nanoparticle then collides with neutral gas which cools it down to $T_{\mathrm{g}}$ on a characteristic time scale $\tau_{\mathrm{cool}}$. Any evaporation has to occur within a window of temperatures between $T^*_{\mathrm{peak}}$ and $T^*_{\mathrm{crit}}$ because, as soon as the temperature has dropped below $T^*_{\mathrm{crit}}$, the nanoparticle is stable. The net probability of growth after the titanium pickup reaction in Eq. (22) is therefore $(1 - P_{\mathrm{evap}})$, where $P_{\mathrm{evap}}$ is the probability of evaporation during the time the nanoparticle is in the evaporation window.

We can now summarize the situation based on figure 3. Each step in the growth from dimers to stable nuclei $\mathrm{Ti}_{N^*}$ (with the stable size $r^*$) is an isolated event. In such an event the nanoparticles start at the gas temperature $T_{\mathrm{g}}$ and get a temperature increase $\Delta T^*_{\mathrm{NP}} = T^*_{\mathrm{crit}}$ when a titanium atom is added, obtaining the temperature $T^*_{\mathrm{peak}} = T^*_{\mathrm{crit}} + T_{\mathrm{g}}$. It is then cooled by collisions with the process gas, with a characteristic time constant $\tau_{\mathrm{cool}}$. Only for a first short time during this cooling process is vaporization possible. We call this the evaporation window. The net probability for the reverse reaction of Eq. (22), $\mathrm{Ti}_{(N+1)} \rightarrow \mathrm{Ti}_N + \mathrm{Ti}$, is given by

$$P_{\mathrm{evap}} = \int k(N, T^*_{\mathrm{NP}}) \mathrm{d}\, t \qquad (27)$$

where $k(N, T^*_{\mathrm{NP}})$ is the evaporation rate [atoms per second], and the integral is evaluated over the time in the evaporation window. The classical form of the evaporation rate, Eq. (24), assumes that the pre-exponential factor is independent of the temperature. The critical temperature effect makes it necessary to add $T^*_{\mathrm{NP}}$ in the argument of the pre-exponential factor $A$:

$$k(N, T^*_{\mathrm{NP}}) = A(N, T^*_{\mathrm{NP}}) \exp\left(-\frac{E_{\mathrm{evap}}}{k_{\mathrm{B}} T^*_{\mathrm{NP}}}\right) \qquad (28)$$

Both factors in Eq. (28) are strongly dependent on $T_{\mathrm{g}}$. The example in figure 4 shows that the exponential factor is very sensitive: an increase in $T_{\mathrm{g}}$ by 300 K here increases the evaporation rate by a factor 100. The pre-exponential factor $A(N, T^*_{\mathrm{NP}})$ is sensitive to $T_{\mathrm{g}}$ for two different reasons. First, the factor $A(N, T^*_{\mathrm{NP}})$ is zero for $T^*_{\mathrm{NP}} < T^*_{\mathrm{crit}}$, and should therefore monotonically approach zero in a range in temperature above $T^*_{\mathrm{crit}}$. Second, from figure 3 it is clear that the time duration of the evaporation process approaches zero when $T_{\mathrm{g}}$ approaches zero. The process analyzed here, addition of single titanium atoms, always puts the nanoparticles in the range just above $T^*_{\mathrm{crit}}$ where these two effects are most important.

---

[2] For such a linear relation between temperature and thermal energy to hold, $\vartheta$ must not vary too much with temperature. We approximate the heat capacity to be temperature-independent. It has been shown that the vibrational heat capacity per atom for $Au_3$ particles are at most 1/3 of that of bulk Au at 300 K. This deviation from bulk decreases with increasing temperature or increasing size [41].

In summary regarding process number 7 in figure 2, Ti atom loss: in the evaporation rate of Eq (28) both the exponential factor and the pre-exponential factor are, for separate reason, steep functions of $T_{\mathrm{g}}$. Together, they make $T_{\mathrm{g}}$ the single most important parameter for the growth from dimers to stable size $r*$.

## 3. Discussion

In this section we will discuss how the process parameters $Q_{\mathrm{Ar}}$, $Q_{\mathrm{He}}$, and $p$ influence the growth environment of the nanoparticles during nucleation, in Zone 2. The framework of the analysis is the process flow chart in figure 2, and in the discussion we will refer to the detailed analysis of the processes in Section 2. The goal is to establish the physical links from the two varied process parameters in the $(p, Q_{\mathrm{Ar}})$ survey of figure 1 (c) to the nucleation limits 1 to 4, also represented by the equations (1) to (4).

### 3.1. Dimer formation: two alternatives

Dimers are formed at some rate $R_{\mathrm{dim}}$ [$\mathrm{m^{-3}s^{-1}}$], and have an average lifetime $\tau_{\mathrm{dim}}$ which is determined by the most efficient process to destroy them: dissociation, or growth to bigger size. If the characteristic time scale $\tau_{\mathrm{c}}$ for variation of the ambient parameters is large in the sense that $\tau_{\mathrm{c}} \gg \tau_{\mathrm{dim}}$, then the dimer density approaches an equilibrium at which production of dimers is in balance with the loss rate,

$$n_{\mathrm{dim}} = R_{\mathrm{dim}} \tau_{\mathrm{dim}} \tag{29}$$

In Section 2.5 the production rates $R_{\mathrm{dim}}$ were estimated for two-body reactions ($Ti^+$ + $H_2O$) giving TiO dimers, and for three-body reactions ($Ti^+$ + Ti + Ar) giving $Ti_2^+$ dimers. Let us start by comparing these two production rates.

#### 3.1.1. The production rates $R_{\mathrm{dim}}$ for TiO and $Ti_2$.

We begin with a numerical example. We consider the situation at the exit of the hollow cathode during the first 150 µs of the pulse when the material ejected has not significantly expanded. Comparing the three-body collision rate $R_{\mathrm{Ti_2^+}}$ of Eq. (21) to the 2 body collision rate $R_{\mathrm{TiO^+}}$ at ultrahigh vacuum, given in the text above Eq. (17), we find that at 200 Pa and a temperature of 1250 K, the number of three-body collisions is of the order of $2.67 \times 10^8$ during the first 150 µs of the pulse. These three-body collisions are about 5 times more frequent than the two-body collisions. Lower $T_{\mathrm{g}}$ and higher $p$ would further increase the three-body collisions compared to the two body collisions. By dividing $R_{\mathrm{Ti_2^+}}$ with $R_{\mathrm{TiO^+}}$, we can broaden the picture to see in what regime of process pressure vs base pressure the different collisions are dominating.

They are equally important when

$$\frac{R_{\mathrm{Ti_2^+}}}{R_{\mathrm{TiO^+}}} = \frac{n_{\mathrm{Ti^+}} n_{\mathrm{Ti}} b^3 n_{\mathrm{Ar}} v_{\mathrm{relA}} \sigma_{\mathrm{Ar-Ti^+}}}{\sigma_{\mathrm{Ti^+\,H_2O}} v_{\mathrm{relH}} n_{\mathrm{\,H_2O}} n_{\mathrm{Ti^+}}} = 1\ . \tag{30}$$

This is under the assumption that we compare only dimer-forming collisions that occur within a volume as large as the neutral gas cloud, which expands with the diffusion speed from Eq. (11). This assumption is motivated by that dimers continue to grow to stable size $r^*$ by adding titanium atoms, as discussed in section 2.6. Dimers created outside of the neutral titanium cloud will not therefore have as high density of growth material to sustain their further growth. The limit between the two regimes is plotted in figure 5, under the assumption that $T_\mathrm{g}$ is 1250 K. The time period is from pulse start to 150 µs within a cloud of constant radius of 2.5 mm.

What is found in figure 5 is that in the UHV experimental setup the three-body collision rate at the cathode exit is larger than the two-body collision rate. Higher process pressures greatly favor three-body collisions over 2 body collisions. In the earlier experiments in a high vacuum system [5] the experiments were well within the two-body dominated regime due to the high impurity background of the vacuum system. The error bars in the blue line come from uncertainty in determining the interaction volume of the titanium ion and titanium neutral, $\gamma$ in Eq. (21).

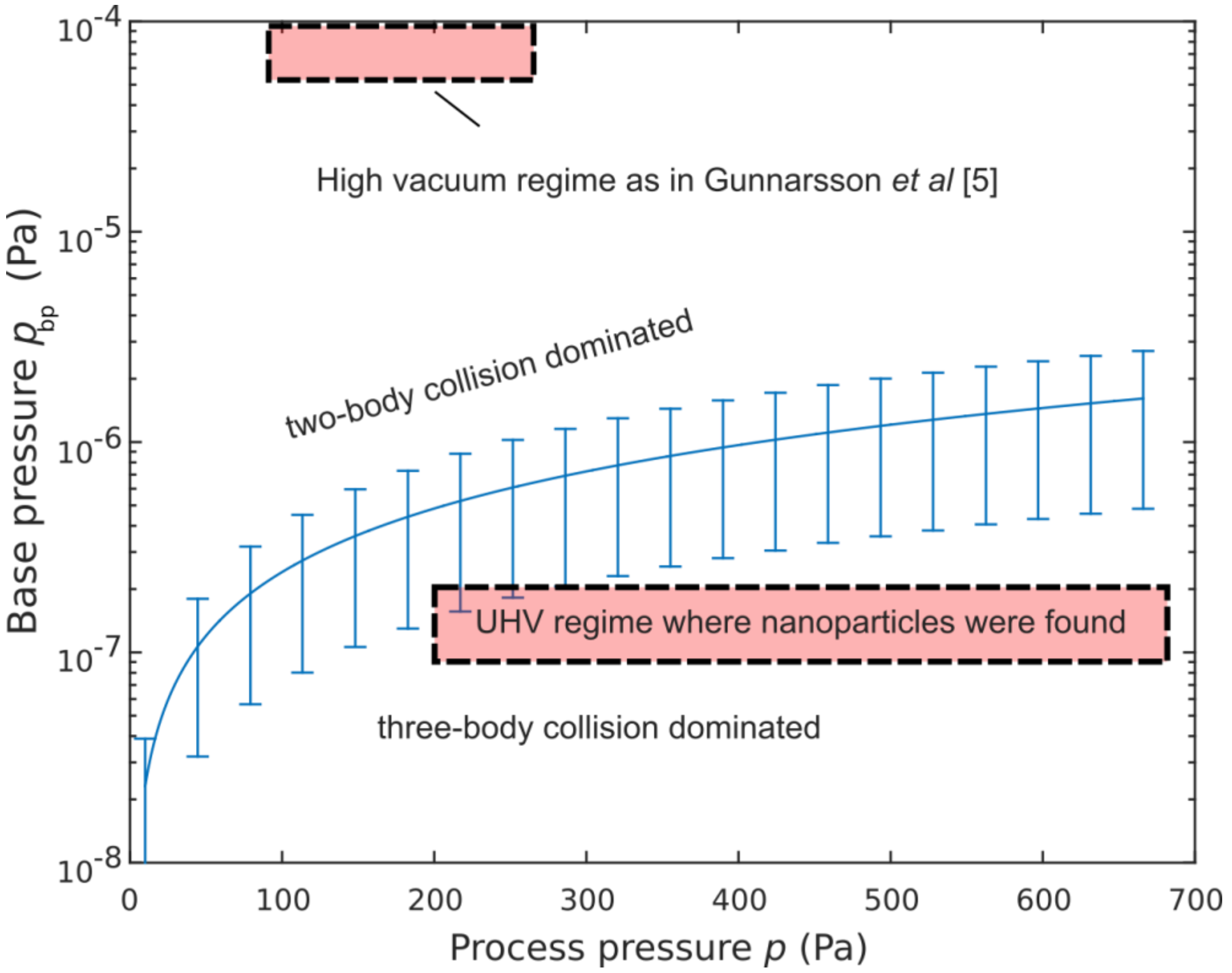

*Figure 5 the separation of the two-body dominated regime with the three-body dominated (blue line) at the exit from the hollow cathode, for typical parameters during the first 150 µs after pulse start. The experiments performed in the UHV system (lower dashed square) is well within the three-body dominated regime. This is compared to the experiments performed in high vacuum* [5]*, which is in the 2 body dominated regime. An increased process pressure greatly favors three-body collisions while an increased base pressure favors 2 body collisions.*

Now let us introduce the time of expansion of the cloud of sputtered material as it moves away from the hollow cathode. Expansion will make the process more two body collision dominated and we are interested to see when this happens. We assume that the cloud has expanded to the same radius as the hollow cathode and investigate how long time it takes for the collisions within the puff to be dominated by two body collisions. The non-process parameter influenced constants in Eq (30) are here combined to a constant C which makes it possible to get an overview of how the boundary between the two regimes depends on the internal process parameters:

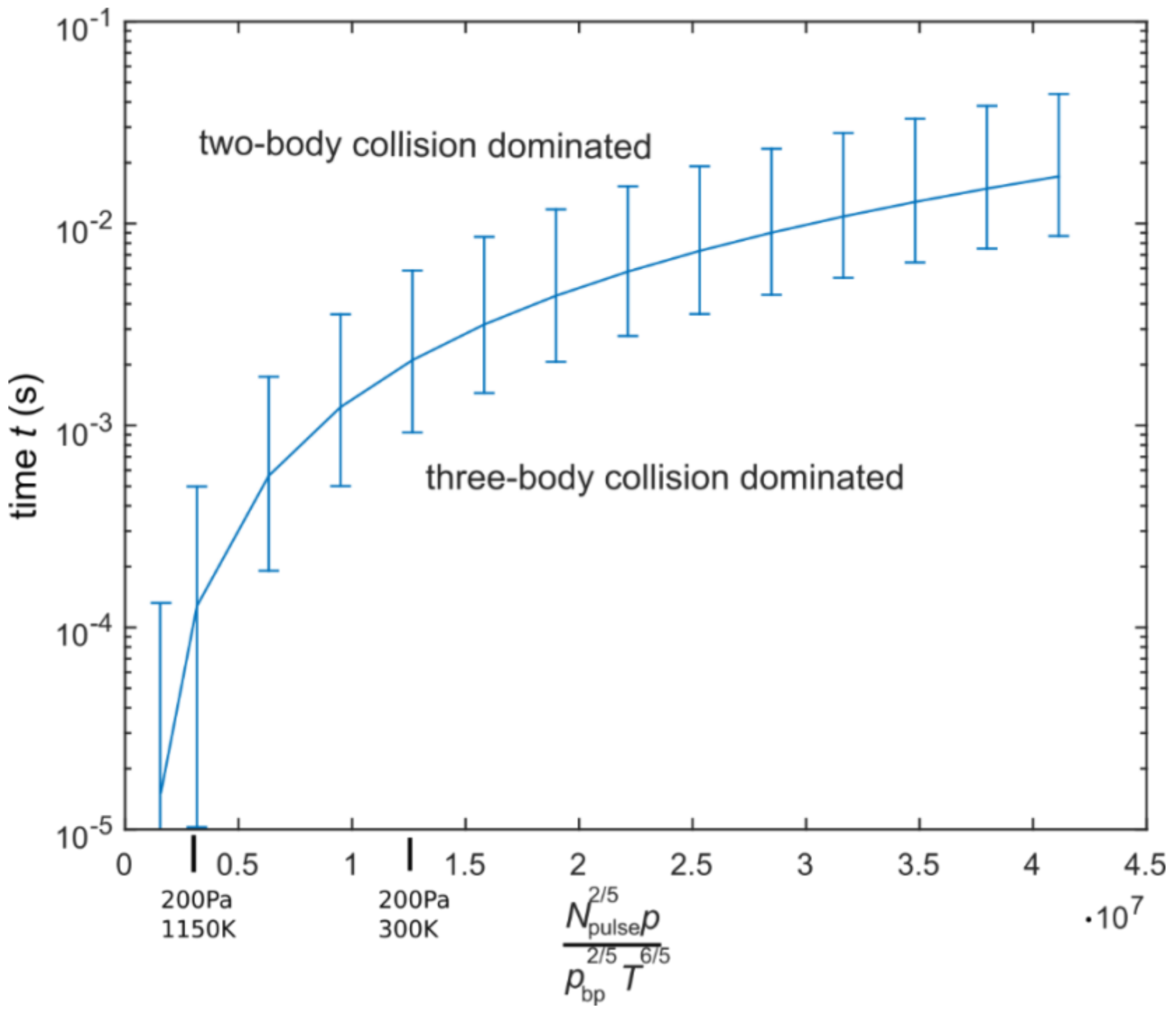

*Figure 6 Y-axis: The time, after ejection from the hollow cathode, for which the expanding neutral gas cloud is dominated by 3- body collisions (below blue line) or 2 body collisions (above blue line). The variable for the X-axis is a normalizing combination of the process parameters. The error bars are from the uncertainties of the constant $\gamma$. Two numerical examples are shown: for a base pressure of 1.3x10$^{-7}$ Pa, p = 200 Pa, and a temperature of 300 K it takes 2100 µs to go in to the two-body dominated regime while, and at a temperature of 1150 K, it takes 100 µs.*

$$\frac{R_{Ti_2^+}}{R_{TiO^+}} = C\frac{N_{pulse}p^{5/2}}{T^3 t^{3/2} p_{bp}} = 1\ , \qquad (31)$$

where $p_{bp}$ is the base pressure and $N_{pulse}$ is the amount of neutrals ejected from one pulse. With this relationship, we can get an estimate at which time periods after the pulse start that the process is dominated by two- or three-body collisions. These times can be approximately translated to a distance from the hollow cathode, by using an approximate ejection speed of 100 m/s from Hasan *et al* [12]. From the example in figure 6 we see that the temperature is the most dominating factor determining whether the process is of a 2- or 3-body nature. At a gas temperature of 1150 K, and a process pressure of 200 Pa, the process becomes dominated by 2 body collisions after 100 μs. however, if the gas temperature is 512 K, it takes 666 μs for the puff of neutrals to transition to a two-body dominated regime. The time of 666 μs is the time between 2 pulses in the current experimental setup, which means that at temperatures below 512 K, there will always be a three-body dominated regime in the experimental setup.

### 3.1.2. The lifetimes $\tau_{\mathbf{dimer}}$ for TiO and $Ti_2$.

The important question is which type of dimer formation that gives the highest density $n_{\mathrm{dim}}$. As can be seen from Eq. (29), this involves both and production rate $R_{\mathrm{dim}}$ and the lifetime $\tau_{\mathrm{dim}}$. The mechanisms that determine the lifetimes of the two types of dimers are very different. For the TiO dimer we can base the discussion on the binding energies in Table I, which refers to reactions in a Ti – $O_2$ gas mix[3]. There are two steps in the dimer formation,

$$\mathrm{Ti} + \mathrm{O}_2 \rightarrow \mathrm{TiO}_2^* \rightarrow \mathrm{TiO} + \mathrm{O} \tag{32}$$

The first step in the reaction releases 7.87 eV, and the second step requires only 6.01 eV. Therefore, the excited molecule $\mathrm{TiO}_2^*$ immediately splits up. This leaves an internal energy of at most (depending on how much the O atom carries away) 1.86 eV in the newly created TiO dimer. Dissociation of TiO requires 7.15 eV, and therefore this dimer is strongly bound right from the start. Its lifetime is determined by processes that destroy TiO. Possibilities for destruction are *e g* collisions with energetic electrons, collisions with metastable Ar* atoms, and further growth either by picking up a titanium atom, or adding oxygen in a two-step reaction of the same type as in Eq. (32). The $Ti_2$ dimer is weakly bound, only 0.833 eV, and it is created in the three-body process between neutrals which leaves it in an excited state just below dissociation. Dissociation is therefore here much more likely through all the types of collisions above. Furthermore, due to the low binding energy, collisions with gas atoms at 1000-2000 K gas temperature can cause significant dissociation. Ionized $Ti_2$ dimers are however more strongly bound with an energy of 2.16 eV. This decreases their likelihood of dissociation from collisions with the gas, but increases their likelihood of dissociation by recombining with electrons.

A quantitative comparison between the lifetimes of the two types of dimers is outside the scope of this paper. We only note that it is very likely that they have different lifetimes and that $\tau_{\mathrm{TiO}} \gg \tau_{\mathrm{Ti}_2}$. Referring to Eq. (29), this has the effect to counteract the difference in the reaction rates $R_{\mathrm{TiO}}$ and $R_{\mathrm{Ti}_2}$. As regards to which reaction dominates the dimer production in the UHV system we can therefore only draw the following two limited conclusions. First, there is a possibility that the uncertain density of contaminant $H_2O$ in zone 2 is so high that the first dimers are mainly TiO. Second, if this is not the case, the three-body reaction can give $Ti_2$ dimers at a sufficient rate to explain the observed production of nanoparticles.

---

[3] These DFT calculations were made for a Ti-$O_2$ mix for the reason that titanium interactions with intentionally added $O_2$ is of broader general interest than interactions with contaminant $H_2O$. We propose that the general conclusions, as regards binding energy and routes to growth, should be essentially the same if the oxygen came from $H_2O$.

Independent of how the first dimers are formed, however, their rate of formation is always higher when the density of growth material is higher. This gives a pressure effect that is very strong for the three-body reaction: Eq. (21) shows that $R_{\mathrm{Ti}_2} \propto p^4$. From the process flow chart of figure 2, the main reason can be identified: a lowered pressure gives lower $n_{\mathrm{g}}$, and in addition lowers the density of growth material which becomes faster diluted by diffusion. For this reason we propose that the limit number 3 in figure 1 (c), the pressure limit, is associated with the dimer formation.

### 3.2. The growth from dimers to stable size $r^*$

In classical nanoparticle growth theory [23], the size limit between growth or shrinking (the stable size $r^*$) is determined by the question if there is a net flux of metal atoms to or from a nanoparticle. If the nanoparticle ensemble, in the whole size range from atoms up to $r^*$, is in thermal equilibrium with a surrounding metal vapor, then this can be treated as a thermodynamic problem. This gives directly, without considering the fluxes, a minimum stable size $r^*$ where further growth reduces the Gibbs free energy G of the system (solid + gas phase). The probability of growth by addition of atoms to this size is then obtained by the density of the atoms multiplied by a Bolzmann factor $e^{-(\frac{\Delta G}{k_{\mathrm{B}}T})}$. In our case this approach is not a possible for two reasons. The first reason is the single-event nature of the evaporation process as illustrated in figure 3. For short times, of the order of 0.1 µs after the addition of a titanium atom, the nanoparticle is much hotter than the gas. This is very far from thermal equilibrium, and the evaporation has to be evaluated as a function of time for each event. The second reason is the pulsed nature of the process. Due to the fast temporal variation in the density of the growth material in a pulse, the relative densities of nanoparticles of different size do not have time to establish equilibrium. This situation calls for a different approach to the growth from dimers to the stable size[4] $r^*$. We will begin with the UHV case, where oxygen is not involved.

#### 3.2.1. Growth from dimers to $r^*$ without oxygen

In each step in the growth of pure Ti nanoparticles from dimers to the stable size $r^*$ there is a balance between the addition of a titanium atom, through collisions plus sticking,

$$\mathrm{Ti}_N + \mathrm{Ti} \rightarrow \mathrm{Ti}_{(N+1)} \tag{33}$$

and the reverse reaction by evaporation,

$$\mathrm{Ti}_{(N+1)} \rightarrow \mathrm{Ti}_N + \mathrm{Ti}. \tag{34}$$

In section 2.7 we demonstrated that the evaporation rate in our type of process is an extremely steep function of the gas temperature. In our pulsed plasma device, the gas temperature has strong gradients in space and also varies rapidly in time. The nucleation, *i.e.* the growth from dimers to stable size, is in this situation assumed to be mainly determined by the variations in the evaporation rate of Eq. (34) due to the gas temperature. We therefore, here, disregard variations of the sticking coefficient which influence the Ti addition rate of Eq. (33).

[4] We here define $r^*$ as the size at which the addition of a Ti atom gives a nanoparticle with a 50% probability of evaporation, during the time that its temperature is in the evaporation window illustrated in figure 3. This gives a 50% probability to further growth.

The nucleation rate is determined by the product of the growth probabilities for all steps from dimers to stable nanoparticles with $N = N^*$. The key to these growth probabilities is the probability of evaporation after the addition of a titanium atom. This problem was analyzed in Section 2.7 above, where it is shown that the evaporation process has a single-event nature. Each addition of a titanium atom heats up the nanoparticle, after which it is rapidly cooled down to the gas temperature as shown in figure 3. The probability of atom loss during this process is a function of both the nanoparticle size and the temperatures, $k(N, T_{\mathrm{NP}}^*)$, which is given in Eq. (28). This equation contains an exponential factor and a pre-exponential factor which both, for separate reason, are very steep functions of $T_{\mathrm{g}}$. The result is therefore probably an almost step-wise effect, such that growth to $r^*$ becomes significant only when $T_{\mathrm{g}}$ is below some critical value. $T_{\mathrm{g}}$ is consequently the single most important parameter for the growth from dimers to stable size $r^*$. This gives a situation with two opposing mechanisms: with increased distance from the hollow cathode, the titanium density decreases (disfavoring nucleation) and $T_{\mathrm{g}}$ decreases (favoring nucleation). Even if dimers are formed at high rate, the further growth to $r^*$can only occur if $T_{\mathrm{g}}$ has dropped far enough.

Combining the observations above, we propose the following explanation to the argon flow limit marked (2) in figure 1 (c). The dimer formation rate is assumed to be sufficient to the right of the pressure limit marked (3). The question therefore is if $T_{\mathrm{g}}$ is too high for the further growth to $r^*$. As shown in Section 2.1, the gas flow $Q_{\mathrm{Ar}}$ determines the extent of the hot zone 2, and thereby how rapidly $T_{\mathrm{g}}$ decreases with distance from the hollow cathode. Above the $Q_{\mathrm{Ar}}$ limit (2), the gas temperature $T_{\mathrm{g}}$ is proposed to be too high in the volume close to the hollow cathode, in which the titanium density is high enough for a significant growth from dimers to $r^*$.

### 3.2.2. Oxygen-assisted growth to $r^*$

In an oxygen-rich environment, such as in the high vacuum system, the situation is different for two reasons. First, the growth by oxidation generally results in a nucleus that is stable and, second, titanium atoms are stronger bound in oxidized nanoparticles which reduces their evaporation rate at a given temperature. Let us look closer at these two effects.

In the growth from dimers to $r^*$, the short vertical red arrows in Table 1 symbolizes two-step reactions of the same type as the dimer formation in Eq. (32) above,

$$\mathrm{Ti_xO_y} + \mathrm{O_2} \rightarrow \mathrm{Ti_xO^*_{(y+2)}} \rightarrow \mathrm{Ti_xO_{(y+1)}} + \mathrm{O}, \qquad (35)$$

where the first step releases enough energy to kick out an oxygen atom. Although such reactions release net energy, and therefore heat up the nanoparticle, this energy is in all cases in Table 1 lower than the energy needed for the evaporation of a titanium atom. In terms of the discussion in Section 2.7, these nanoparticles are therefore below the critical temperature, and their vapor pressure is zero. These steps in the arrow-marked growth path in Table 1 are therefore safe from evaporation.

Let us now compare the binding energies of pure titanium nanoparticles (the first row) with the binding energy of oxide nanoparticles growing along the red arrows, which denotes the growth path for the highest binding energy in every step. We see that the average binding energy along the oxide route is 5.66 ± 1.57 eV, much higher than the average for pure titanium nanoparticles, which is 2.48 ± 0.66 eV. A higher binding energy is also true for the $Ti_8O_8$ and the $Ti_6O_{10}$ nanoparticles which shows that the trend holds for larger nanoparticles as well. In addition to this, we can see from tabulated data that the vapor pressure of bulk titanium monoxide is 0.2832 Pa at 1998 K [24] and 0.952 at 1998 K [25] for pure titanium. Thus it is reasonable to conclude that nanoparticles containing oxygen are more stable at higher temperatures than pure titanium nanoparticles. The oxygen content within the nanoparticles has to be large enough for this to have an effect. This is because if there is only enough oxygen to create a dimer, it does not decrease the growing particles' vapor pressure, and thus does not help it reach $r^*$. This conclusion is consistent with the observation in [5] where only particles with a significant oxidation was possible to be synthesized at that temperature. To see if growth by adding oxygen is possible also for ionized nanoparticles, additional calculations were performed which are presented in Table 2, where it can be seen is that the same general trend holds for ionized nanoparticles as well.

We thus conclude that titanium in nanoparticles that grow in an oxygen rich environment is stronger bound than titanium in nanoparticles that grow in an oxygen starved environment. It therefore is crucial to use process conditions that lead to a low gas temperature in the growth zone if non-oxidized titanium nanoparticles should be able to grow.

In summary: the growth to $r^*$ is possible at higher gas temperature if the nanoparticle is kept significantly oxidized during this growth process. The immediate result of the larger binding energy, in an oxygen-rich environment, is therefore a much larger process window for nanoparticle growth as shown in figure 2 (c). As shown by Gunnarsson *et al* [5], the limit to this process window, the oxygen limit marked (1), corresponds to a required lowest density of oxygen-containing contamination. Unfortunately, this level was found to be so high that also the final size nanoparticles became significantly oxidized.

### 3.3. The role of He

The role of helium in the discharge is interesting. Adding helium is necessary because, without helium, no nanoparticles are produced [4]. However, also argon is needed, since a discharge in pure helium gives no nanoparticles without oxygen. A mix of the two gases is consequently needed, but this also causes problems in the form of a discharge instability which limits the process window [4].

The main role of helium is proposed to support the growth from dimers to stable size $r^*$ by keeping the temperature of the nanoparticles low. However, we find here that this is an indirect effect, not the direct cooling by helium atoms onto the nanoparticles. As discussed in section 2.3, this direct cooling rate increase only by approximately 31 % if the argon atoms are replaced by helium. The main effect of helium is instead that it increases the thermal conductivity in zone 3 of figure 1 (a), between the walls and the zone 2, in which the nucleation occurs. There is a 775 % increase of the thermal conductivity of the gas for the same substitution, argon to helium. We can thus conclude that the primary purpose of using helium gas is not to cool the nanoparticle directly but rather to aid in the cooling of the hot vapor ejected from the cathode. This gas temperature, in turn, influences the probability of growth to a stable size $r*$ through a "more than exponential" effect on the evaporation rate, as discussed in Section 2.7.

**4 Summary and conclusions**

A model is presented for experimentally obtained nucleation of pure titanium nanoparticles, from sputtered titanium vapor, in an ultra-high vacuum system. In this experiment, a high partial pressure of helium is added to the process gas, instead of earlier used trace amounts of oxygen. The effect of helium is that it cools the process gas in the region where nucleation occurs, which is important for two reasons. First, a reduced gas temperature enhances $Ti_2$ dimer formation, which is proposed to be dominated by three-body collisions between titanium ions, titanium neutrals and argon atoms (helium is not injected through the hollow cathode, and therefore disregarded for this first step in the nucleation process). This reaction is enhanced mainly because a lower gas temperature gives a higher gas density at a given pressure, which in turn reduces the dilution of the titanium vapor through diffusion. This first effect can, therefore, also be achieved by increasing the gas pressure. Second, a reduced gas temperature has a "more than exponential" effect in lowering the evaporation rate of the nanoparticles during their further growth from dimers to stable nuclei of size $r*$, the second step in nucleation. It is shown that this evaporation mechanism is not possible to model thermodynamically, as is often done. Instead, a single-event nature of the evaporation process has to be considered. This leads to an evaporation probability from nanoparticles that is zero below a size-dependent critical nanoparticle energy-equivalent temperature $T_{\mathrm{NP}}^*$, which is the reason for the phrase "more than exponential". Together, the mechanisms above explain two experimentally found limits for nucleation in an oxygen-starved environment, see figure 1 (c). First, there is a lower limit to the pressure which is related to the first step in nucleation, dimer formation. Second, above this pressure limit, there is an upper limit to the gas temperature above which evaporation makes the further growth to stable nuclei impossible. These findings are schematically summarized in figure 7. The argon gas flow will influence the gas temperature within the zone near the hollow cathode into which the sputtered titanium ions and neutrals are ejected. A high temperature increases the diffusion rate of the sputtered material, which decreases the amount of dimers formed. The further growth of the nanoparticles to a thermodynamically stable size is hindered if the gas temperature is too high. This is because the evaporation window in energy-equivalent temperature $T_{\mathrm{NP}}^*$ (see section 2.7) is as wide as the gas temperature and because the evaporation rate has an exponential temperature dependence within this window.

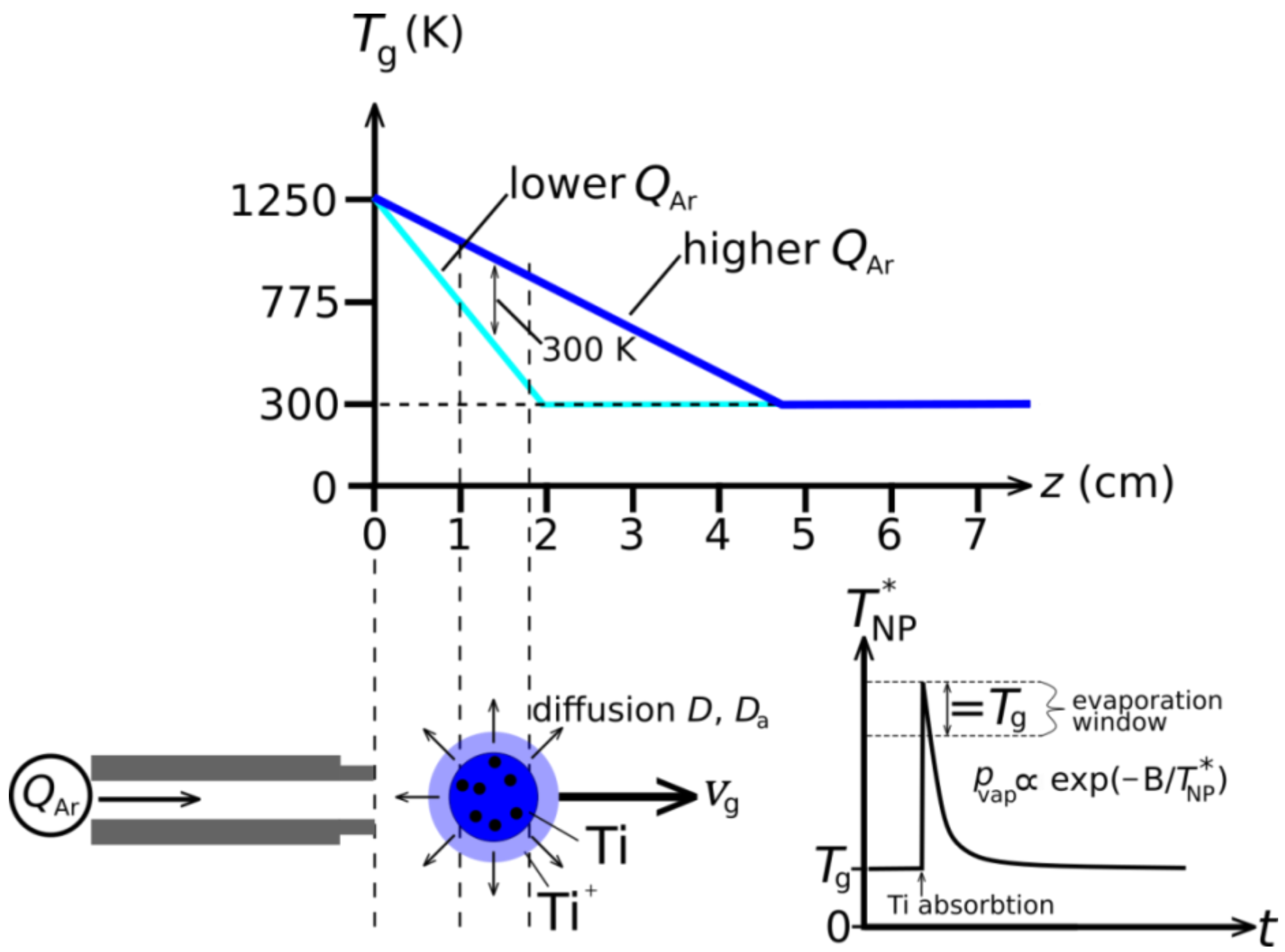

*Figure 7 Schematic summary of the gas flow's influence on the nucleation process. The temperature zone will extend further out from the hollow cathode at higher argon gas flows (top). This will increase the diffusion of the titanium ions and neutrals ejected from the hollow cathode (bottom left). Also, the nanoparticles' growth is counteracted by evaporation (bottom right), and for this the gas temperature has two effects. First, the nanoparticle evaporation window is as wide as the gas temperature and, second, their evaporation rate within this window has an exponential temperature dependence.*

Some experiments at a high base pressure, with significant process gas contaminations, were also analyzed. Here, the collisions creating the first dimers are two-body collisions between a titanium ion and a water molecule. The binding energies of oxidized titanium nanoparticles are calculated by density functional theory. It is found that titanium is generally stronger bound in oxidized than in pure titanium nanoparticles, giving an "easy oxygen route" to a stable size $r*$. This route, however, seems to require so high oxygen densities that also the final nanoparticles become oxidized.

In this work, most experiments were made in the form of $(Q_{\mathrm{Ar}}, p)$ surveys at a constant $Q_{\mathrm{He}}$ of 55 sccm, and at constant pulse parameters (pulse length, frequency, and current). Consequently, there is considerable room for experimental optimization of the parameter combinations of pressure, argon flow, helium flow, and pulse parameters.

**Acknowledgements**

This work was made possible by financial supported by the Knut and Alice Wallenberg foundation (KAW 2014.0276) and the Swedish Research Council under Grant No. 2008-6572 via the Linköping Linneaus Environment - LiLi-NFM. We also acknowledges financial support from the Swedish Government Strategic Research Area in Materials Science on Functional Materials at Linköping University (Faculty Grant SFO Mat LiU No 2009 00971) and from the Swedish Research Council (VR) Grant No. 2016-05137_4.

## Appendices

### Appendix 1: Details on three-body reactions

The expression for the rate for the three-body collision is adapted from Smirnov [19], by using the collision cross section and Lennard-Jones potential for ions instead of neutrals to fit the highly ionized plasma in the current experimental setup

$$R_{\mathrm{Ti^+TiAr}} = n_{\mathrm{Ti^+}} \cdot n_{\mathrm{Ti}} \cdot n_{\mathrm{Ar}} \cdot v_{\mathrm{relAr}} \cdot b^3 \cdot \sigma_{\mathrm{Ti^+Ar}}, \qquad \text{(A1)}$$

where $n_{\mathrm{Ti}}$ is the density of titanium, $n_{\mathrm{Ti^+}}$ is the density of titanium ions, $n_{\mathrm{Ar}}$ is the density of argon atoms, $v_{\mathrm{relAr}}$ is the relative collision velocity between an argon atom and a titanium atom, $\sigma_{\mathrm{Ti^+Ar}}$ is the cross section for collisions between argon neutrals and the titanium ion and $b$ is the critical radius for interaction between titanium ion and titanium neutral. There are three reasons that one ion has to be involved in the three-body process. Firstly, the ion will have a larger collision cross section with the argon atom due to an induced dipole interaction. Secondly, there are more titanium ions in the gas compared to titanium neutrals [12]. Lastly the interaction volume $b^3$ is larger between a titanium ion and a titanium neutral due to the attractive inverse forth power term [26]. The distance $b$ was estimated to be the distance at which the Lennard-Jones potential is equal to the thermal energy [19]. Since we are interested in the interaction between ions and neutrals we use the 12-6-4 Lennard-Jones potential [27].

$$U(b) = 2\epsilon\left[(1+\gamma)\left(\frac{R_0}{b}\right)^{12} - 2\gamma\left(\frac{R_0}{b}\right)^{6} - 3(1-\gamma)\left(\frac{R_0}{b}\right)^{4}\right] = k_B T_g \qquad \text{(A2)}$$

where $\epsilon$ is the binding energy of the ion and neutral, $\gamma$ is a constant approximated to be 0.5 ± 0.4, $R_0$ is the van der Waals radius and $T_g$ is the gas temperature. An approximation is made that it is only the long-range ion interaction in equation 3 that is the most important factor, and thus we can estimate the factors that determine the interaction volume as:

$$b \approx \sqrt[4]{\frac{{R_0}^4 6\epsilon(1-\gamma)}{k_B T_g}} \qquad \text{(A3)}$$

The cross section for collision between an ion and a neutral is given by the Langevin capture cross section [28]

$$\sigma_{\text{ion-neutral}} = \pi\sqrt{\frac{\alpha_{Ar} {q_i}^2}{\epsilon_0 8 k_B T_g}} \qquad \text{(A4)}$$

where $\alpha_{Ar}$ is the polarizability of the neutral, $q_i$ is the charge of the titanium ion, and $\epsilon_0$ is the permittivity of vacuum.

**Appendix 2: Why adding charged particles gives no net growth in the nucleation phase**

A possibility to growth is by adding charged particles, giving the net reaction

$$\text{Ti}_N + \text{Ti}^+ + e \rightarrow \text{Ti}_{(N+1)}. \qquad \text{(A5)}$$

Please note, that the ion and the electron in the reaction of Eq. A5 are in reality added individually; the electron can come before the ion, or the other way around. Whichever is the case, the net effect reaction A5 is that the ionization energy $E_{i,Ti}$ of Ti has to go somewhere. The total amount of energy released in reaction A5 is easily determined by a thought experiment: first, let $\text{Ti}^+$ and $e$ in the reaction of Eq. A5 recombine, before being added to the nanoparticle. This releases the amount $E_{i,Ti}$ of energy. Then, let the created Ti atom be added to a nanoparticle $\text{Ti}_N$. This releases the binding energy $E_{vap}$, giving a total released energy $E_{vap} + E_{i,Ti}$. An uncertain parameter is how much of this released energy that is converted to heat. Fortunately, we do not have to answer this question here. It is sufficient to note that there is more energy released in the addition of charged particles ($E_{vap} + E_{i,Ti}$) than in the addition of atoms ($E_{vap}$). In section 2.7 we show that for small nanoparticles, as considered during the growth from dimers to $r*$, the rate of evaporation in the reaction $\text{Ti}_{(N+1)} \rightarrow \text{Ti}_N + \text{Ti}$ increases dramatically with the nanoparticle temperature. For this reason we propose that the addition of charged particles, the reaction of Eq. (A5), should usually be followed by evaporation of a titanium atom. Addition of $Ti^+$ ions then gives no net growth during the nucleation phase.

**Appendix 3: Binding energies of atoms to nanoparticles**

Quantum-chemical computations were carried out in order to obtain binding energies in small titanium and titanium oxide nanoparticles, which compose our models for the nanoparticles. Hybrid density functional theory (DFT) employing the B3LYP functional [29] [30] and the basis set 6-311++G(2d,2p) [31] [32] [33] were used as implemented in the Gaussian09 [34] program. The geometries of the nanoparticles were optimized by minimizing their electronic energies with respect to the nuclear coordinates using the aforementioned program. The initial geometries were obtained largely by trial-and-error, either starting from regular geometric shapes (with atoms added or subtracted from the corners or facets) or from reported geometries for similar nanoparticles in the literature [35] [36] [37] [38]. For the titanium oxide nanoparticles, trial initial structures were also obtained by extracting coordinates for neighboring atoms from the anatase, rutile ($TiO_2$), corundum ($Ti_2O_3$) or rock salt (TiO) crystal structures. The reported energy for each species correspond to the spin state and geometry with the lowest electronic energy found.

The binding energies for adding a titanium atom were obtained from the difference between the electronic energy of the original nanoparticles and the sum of the electronic energy of the nanoparticles with the Ti atom added and the electronic energy of a Ti atom. The results are presented in Table 1 and Table 2.

It can be mentioned that the computed energies for the $Ti_2O_4$ and $Ti_5O_{10}$ nanoparticles agree well with the stoichiometric $TiO_2$ molecules reported in Ref. [39], were the same functional and a similar basis set were used, i.e. the normalized nanoparticles energy as defined and reported in that reference is for the first nanoparticles 57.3 kcal/mol whereas we get 57.2 kcal/mol, and for the second nanoparticles 92.9 kcal/mol versus 91.4 kcal/mol.

**Appendix 4. Uncertainties in the estimate of the contamination of water.**

The base pressure of the vacuum system is measured when there is no restriction on the vacuum pump. To achieve the desired pressure at a given argon gas flow, the pump is restricted with a baffle. This will increase the partial pressure of contaminants such as water vapor in the vacuum system [40], since less water vapor can be pumped away. But on the other hand, sputtering a titanium cathode will decrease this vapor pressure due to gettering on the titanium coated vacuum chamber wall. Experiments with a residual gas analyzer (not presented) show that this titanium hollow cathode can decrease the partial pressure of externally supplied oxygen to the growth zone by as much as $3.8 \times 10^{-2}$ Pa. It is thus un-certain if the partial pressure of contaminants within zones 2 and 3 is higher or lower than the base pressure. We thus approximate it to be the same.